\documentclass[manuscript,screen]{acmart}

\usepackage{graphicx}
\usepackage{amsmath, amsfonts,amssymb,epsfig, epstopdf, subfigure}
\usepackage{color}
\usepackage{array}
\usepackage{multirow}
\usepackage[utf8]{inputenc}
\usepackage[english]{babel}
\usepackage{amsthm}
\usepackage{graphicx}
\usepackage{epsfig}
\usepackage{color}
\usepackage{subfigure}
\usepackage{verbatim}
\usepackage{url}
\usepackage{algorithm}
\usepackage{algorithmic}
\usepackage{amssymb}
\usepackage{latexsym}
\usepackage{amsmath}
\usepackage{mathtools}
\usepackage{array}
\usepackage{url}
\usepackage{soul}

\AtBeginDocument{%
  \providecommand\BibTeX{{%
    \normalfont B\kern-0.5em{\scshape i\kern-0.25em b}\kern-0.8em\TeX}}}

\newcommand\raisepunct[1]{\,\mathpunct{\raisebox{0.5ex}{#1}}}
\begin{document}


\title{A Prospect Theoretic Approach for Trust Management in IoT Networks under Manipulation Attacks}

\author{Mehrdad Salimitari}
\email{mehrdad@cs.ucf.edu}
\orcid{0001-9243-9480}
\affiliation{%
  \institution{University of Central Florida}
  \city{Orlando}
  \state{Florida}
  \postcode{32816}
}

\author{Shameek Bhattacharjee}
\email{shameek.bhattacharjee@wmich.edu}
\affiliation{%
  \institution{Western Michigan University}
  \city{Kalamazoo}
  \state{Michigan}
  \postcode{40998-5466}
}

\author{Mainak Chatterjee}
\email{mainak@cs.ucf.edu}
\affiliation{%
  \institution{University of Central Florida}
  \city{Orlando}
  \state{Florida}
  \postcode{32816}
}

\author{Yaser P. Fallah}
\email{yaser.fallah@ucf.edu}
\affiliation{%
  \institution{University of Central Florida}
  \city{Orlando}
  \state{Florida}
  \postcode{32816}
}

\renewcommand{\shortauthors}{Salimitari et al.}


\begin{abstract}
As Internet of Things (IoT) and Cyber-Physical systems become more ubiquitous in our daily lives, it necessitates the capability to measure the trustworthiness of the aggregate data from such systems to make fair decisions. However, the interpretation of trustworthiness is contextual and varies according to the risk tolerance attitude of the concerned application. In addition, there exist varying levels of uncertainty associated with an evidence upon which a trust model is built. Hence, the data integrity scoring mechanisms require some provisions to adapt to different risk attitudes and uncertainties. 

In this paper, we propose a prospect theoretic framework for data integrity scoring that quantifies the trustworthiness of the collected data from IoT devices in the presence of adversaries who try to manipulate the data. 
In our proposed method, we consider an imperfect anomaly monitoring mechanism that tracks the transmitted data from each device and classifies the outcome (trustworthiness of data) as {\em not compromised,} {\em compromised,} or {\em undecided}. These outcomes are conceptualized as a multinomial hypothesis of a Bayesian inference model with three parameters. These parameters are then used for calculating a utility value via prospect theory to evaluate the reliability of the aggregate data at an IoT hub. In addition, to take into account different risk attitudes, we propose two types of fusion rule at IoT hub-- optimistic and conservative. 

Furthermore, we put forward asymmetric weighted moving average (AWMA) scheme to measure the trustworthiness of aggregate data in presence of On-Off attacks. The proposed framework is validated using extensive simulation experiments for both uniform and On-Off attacks. We show how trust scores vary under a variety of system factors like attack magnitude and inaccurate detection. In addition, we measure the trustworthiness of the aggregate data using the well-known expected utility theory and compare the results with that obtained by prospect theory. The simulation results reveal that prospect theory quantifies trustworthiness of the aggregate data better than expected utility theory.
\end{abstract}


\maketitle

\section{Introduction}


The proliferation of Internet of Things (IoT) is witnessing an exponential growth, both in terms of market value and number of devices. In situations where multiple devices in a network contend for resources, it is very common that some may deviate from the mutually-agreed upon norms to either i) illegitimately draw additional benefits or ii) mislead a central entity (i.e., IoT hub)  from reaching a fair decision~\cite{bhattacharjee2017preserving}. Thus, there exists a demand for a mechanism that evaluate the trustworthiness of the devices through trust and reputation scores~\cite{yaqoob2017internet,nitti2017trustworthiness}. Usually, trustworthiness is assigned to individual IoT devices based on the quality of information they share with others. Unlike what is claimed in~\cite{nitti}, we argue that the devices may not be always malicious by themselves. It is plausible that only devices inputs are compromised 
by an adversary before they reach the central decision maker (performing a fused decision)~\cite{2017trust}. For example, a man in the middle attack (MITM) that modifies the data sent from devices does not represent a compromised IoT device~\cite{andrea2015internet}. In such a case, focusing on trustworthiness of a device is not appropriate. Rather, the trustworthiness of the {\em aggregate data} at the controlling IoT hub is of the utmost importance. It may be noted that an IoT hub may have a generic fusion rule for the aggregate data. Therefore, the overall trustworthiness depends on both the type of fusion rule and the reliability of the data collected from each device~\cite{singh2014survey}.


Usually, trustworthiness of the input data from different devices in IoT networks is calculated based on the evidence provided by a feedback or an anomaly monitoring system~\cite{pacheco2017enabling}. The anomaly monitoring systems usually evaluate whether an input from an IoT device is satisfactory (denoted as 1) or not (denoted as 0) and expressed as a binary value~\cite{nitti}. However, it is known that it might not always be possible for the feedback or anomaly monitoring systems to express interactions from various devices in strict binary values due to inherent wireless channel uncertainties in wireless IoT networks~\cite{Bhat-milcom}. In these cases, the anomaly monitoring mechanism may not have enough information about data inputs of some devices. therefore, the feedback system will produce a ternary evidence, the third inference being `undecided'~\cite{mendoza2015mitigating}. Hence, each data input from different devices is labeled as {\em compromised} (negative evidence), {\em not compromised} (positive evidence), or {\em undecided} (uncertain evidence). There exist many algorithms for anomaly monitoring which can classify data inputs to these labels with different accuracy~\cite{zarpelao2017survey, stiawan2016anomaly}. However, this classification is outside the scope of this paper. The question is, can trust models handle the uncertainty (undecided outcome) associated with different anomaly monitoring algorithms in an adaptive manner?

Nonetheless, this problem can be further exacerbated because the adversary's behavior may not always be consistent and it can vary under different circumstances (e.g., the adversaries magnitude of attack in a particular time slot, available power, energy constraints, or the state of other network entities). Therefore, the adversary may behave honestly for a while to gain trust and exploit the weaknesses in the trust management scheme so it can attack later~\cite{chae2014trust}. Thus, an adversary's behavior is often dynamic where they rapidly switch to different modes (On or Off) rendering different behavior under various situations which is known as {\em On-Off attacks}~\cite{caminha2018smart}. An On-Off attack tries to masquerade attacks as temporary unintentional noisy conditions~\cite{mendoza2015mitigating}. In such cases, regular trust management mechanisms either fail to react quickly or allow quick trust improvement when the adversary returns to good behavior.

In addition, different systems have different risk tolerance thresholds. For example, a mission critical system may not operate properly even if there are few compromised data inputs because the associated risks are too high. However, personal networks like smart homes are more risk-tolerant. Therefore, we need a trust management scheme that considers the risk tolerance threshold of systems as a factor in computing the trustworthiness of the aggregate data. This factor can be addressed in the type of fusion rule that we use at the IoT hub. 

Therefore, a robust trust management scheme should consider imperfect anomaly monitoring, address On-Off attacks, and be adaptive to different risk tolerance of different systems. There exist several research works trying to tackle some of these issues which are discussed in Section~\ref{rel-work}. However, to the best of our knowledge, there is not any proposed method that addresses all of the mentioned problems at the same time. 

In this paper, we put forward a Bayesian inference model for monitoring possible anomalies caused by adversaries in an IoT network. This model conceptualizes the outcome of monitoring the devices inputs over time as a multinomial
hypothesis. The adversary can manipulate the data sent from any IoT device in the network to the IoT hub.
We assume each IoT device generates a single input in each time slot, all of which are vulnerable to manipulation attacks. We assume an imperfect anomaly monitoring mechanism that produces varying feedback for each device input over time. The three possible outcome of the monitoring mechanism (compromised, not compromised, and undecided) are the parameters for the multinomial hypothesis of our Bayesian inference model.

Thereafter, we propose a prospect theoretic approach to measure the trustworthiness of the aggregate data at the IoT hub. We use the posterior believes obtained by our Bayesian inference model to calculate a utility value using prospect theory. This utility value reflects the reliability of the aggregate data. Prospect theory is an economics theory proposed by A. Tversky and D. Kahneman to overcome the limitations of expected utility theory for modeling behavioral finance~\cite{prospectweight}. Although, expected utility theory has been used for decades for modeling financial problems, it is not robust for decision-making under risk. Prospect theory has been shown to be the most appropriate theory for decision making under risk for economic problems~\cite{barberis2013thirty}. It can also model people's behavior in risky situations better than expected utility theory~\cite{kahneman2013prospect}. We discuss in Section~\ref{sec-integrity-PT} that the data fusion center (IoT hub) has a similar behavior towards adversaries as people's behavior in decision-making under risk. Thus, prospect theory looks very promising to model the fusion center (IoT hub) to measure the trustworthiness of the aggregate data.

Furthermore, we propose two different models for data fusion at IoT hub-- optimistic
and conservative. Thus, our data integrity scoring model can take into account different risk attitudes of different systems.
The optimistic model is applied to systems where some tolerance for wrong decisions is allowed. However, for a mission critical system where there is almost no room for erroneous decisions, the conservative model should be used. Our prospect theoretic approach for computing the utilities allows us to
differentiate between the way the utility is calculated for
optimistic and conservative systems because of intrinsic features of prospect theory~\cite{sanjab2017prospect}. Thus, we can model both loss averse and risk averse systems. On the contrary, expected utility theory is neither loss averse nor risk averse which makes it unable to differentiate between optimistic and conservative systems~\cite{prospectweight}. To verify this claim, we compare our model with an expected utility theory based model in Section~\ref{sec-comparison}
and show that that the use of 
prospect theory is more apt for data that is at risk. We further investigate this claim by simulations in Section~\ref{sec-ptVSeut-simulation}.

Eventually, we introduce an Asymmetric Weighted Moving Average (AWMA) scheme to extend our proposed data integrity scoring model to accommodate the challenges of On-Off attacks. This scheme is defined based on our prospect theoretic approach in order to quickly respond to abrupt changes in adversary's behavior. In Section~\ref{sec-trustmanage}, we show that our AWMA scheme is able to robustly detect stealthy On-Off attacks as opposed to commonly used trust management systems.

We conduct extensive simulation experiments to verify the accuracy of our proposed method for both non-opportunistic (uniform) and opportunistic (On-Off) attacks. We also show how data integrity scores vary under a variety of system factors like attack intensity and inaccurate detection. We observe that with more inputs compromised, the data integrity scores reduce. Low data integrity scores may also be caused by temporal or initial lack of evidence due to uncertainty associated with imperfect anomaly monitoring.

\section{Related Work}
\label{rel-work}


In a time slotted system, a trust management scheme usually updates individual trust scores over time and is usually accompanied by a recovery scheme to allow the system to negate effects of intermittent noise or errors. Popular examples of trust management schemes are cumulative weighted moving average (CWMA) and exponentially weighted moving average (EWMA)~\cite{li2007trust}. The CWMA assigns equal weights to all individual scores and is useful when the goal is to characterize long-term behavior. EWMA, on the other hand, is a forgetting scheme which designates more weight to recent observations than old ones to reflect recent changes in data integrity. However, we show in Section~\ref{on-off-model} that both of these approaches do not work well under On-Off attacks. 


Apart from CWMA and EWMA, there have been other efforts to address limitations of trust management schemes. Chae et al. proposed a trust management technique to address On-Off attacks~\cite{chae2014trust}. Although their proposed model is flexible and takes into account different risk attitudes, it does not consider an imperfect anomaly monitoring mechanism.
Guo et al. devised a 3-tier cloud hierarchical trust management protocol for IoT networks~\cite{guo2017mobile}. This protocol considers an imperfect monitoring mechanism and verifies effects of different attacks including On-Off attacks (opportunistic attacks). However, they show in their simulation results that their method can properly respond to anomalies in the aggregate data if there is not any On-Off attacks. In the presence of On-Off attacks, the accuracy of their method significantly drops. In addition, they do not differentiate between systems with different risk tolerance. Alshehri et al. proposed a clustering-based methodology for trust management that eliminates the outliers of trust scores~\cite{alshehri2018clustering}. Their method increased the scalability of existing solutions. However, in their model, they did not consider imperfect monitoring or different risk tolerance of different systems.

Therefore, we lack a concrete mathematical framework that quantifies general trustworthiness of the aggregate data by taking into account imperfect anomaly monitoring, On-Off attacks, and different risk attitudes in different systems. We propose a Bayesian inference model which considers an imperfect anomaly monitoring mechanism to capture the posterior belief about trustworthiness of future data sent from each device in an IoT network. In addition, we propound a prospect theoretical model to fuse the data from different devices in order to make a fair decision about reliability of the aggregate data~\cite{lu2017lightweight}.

It should be noticed that prospect theory can model people's behavior in risky situations more accurately than expected utility theory~\cite{ramos2014state}. On the other hand, adversaries usually follow the same behavior as people for decision-making. In this regard, L. Xiao et al. have used prospect theory to model continuous, stealthy attacks on cloud storage devices~\cite{xiao2017cloud}. They claim that prospect theory can model these kinds of attacks more accurately than expected utility theory since these attacks exhibit similar behaviors as human beings. We also delineate in Section~\ref{sec-integrity-PT} that IoT hub behaves similar to human beings in decision-making. Therefore, our proposed model is flexible enough to make a reasonable decision for different systems with different risk tolerance.


Previously, prospect theory has been applied to analyze wireless communications and traffic routing~\cite{li2012prospects, yang2015prospect, gao2010adaptive}. In~\cite{li2012prospects}, a random access game is formulated using prospect theory to study channel access between two subjective end-users in wireless networks. In~\cite{yang2015prospect}, Y. Yang et al. have proposed to use prospect theory for resource allocation in cognitive radio networks and pricing which increases the revenue of service providers in presence of subjective users. The suitability of prospect theory for routing decisions in a stochastic network is investigated in~\cite{gao2010adaptive}. S. Gao et al. have modeled the routing decision using both prospect theory and expected utility theory. Using empirical results, they prove that prospect theory is a more appropriate model because of its inherit features such as risk aversion and loss aversion.




\section{System Model and Assumptions}
\label{sec-model}
We consider a time-slotted system with $N$ IoT devices that
each of them creates only one input (i.e., the vote)  in each time slot.
The nature of the decision is generic; it could be as simple as a binary voting or it could be
a complex decision metric. We use a prospect theoretic approach for decision-making. A centralized IoT hub fuses all votes from each component through the fusion
rule (decision-making approach) to reach a global decision.


\noindent
$\bullet${\bf Adversarial model:}
We assume that all the inputs from each IoT device are exposed to an adversary
whose goal is to disrupt the voting process at the central hub.
The adversary has some predefined attack resources and can
choose to attack different sets of inputs over time and also attack varying number of inputs in
each time slot. However, it maintains the attack on a specific long-term fraction of the inputs which we call it the probability of attack denoted by $P_{attack}$. 
For example,  $P_{attack} = 0.6$ means that the
adversary compromises $60\%$ of the inputs over a large period of time.
Hence, a single observation (over one time slot) is not sufficient
for characterizing the behavior of the adversary. 

\noindent
$\bullet${\bf Imperfect anomaly monitoring:}
We assume that there is an anomaly monitoring or failure detection mechanism in place
that infers whether the input from each device has been compromised or not. Unlike related works~\cite{nitti,dchen-11}, we consider that the monitoring mechanism cannot infer an anomaly  
with certainty. Thus, it classifies the inputs into three categories: i) compromised, ii) not compromised, and iii) undecided.
All three are functions of environmental parameters that may be dynamic over time.
Also, the system transients and noisy environments may increase or decrease temporal uncertainty. 
Hence, the data integrity is computed over time-- a larger time window of observations allows
 a more accurate estimation of the overall data integrity.  

\noindent
$\bullet${\bf Uniformly distributed prior inference:}
Since there is no bias (or available information) over any of the three possible outcomes of the monitoring process, 
we assume that the initial probabilities of all outcomes are equal. Similarly, we assume that the prior probabilities of an input being compromised or not is also uniformly distributed.

\noindent
$\bullet${\bf Probability of detection:}
\begin{figure}[!b]
\begin{center}
\includegraphics[width=3.5in, keepaspectratio]{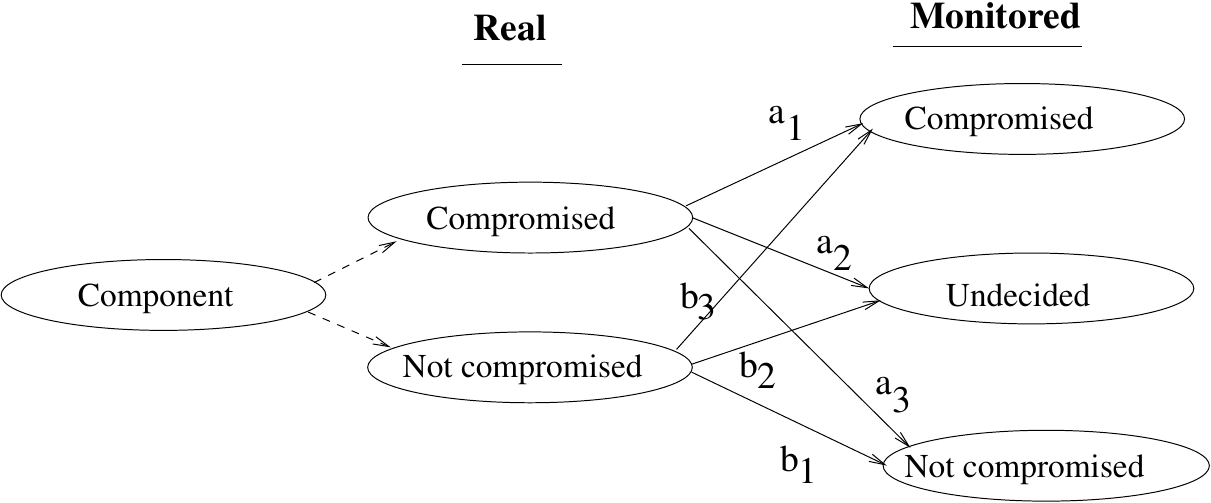}
\caption{Inference possibilities for probability of detection.}
\label{pdet}
\end{center}
\end{figure}

We define the probability of detection as the percentage of IoTs' inputs that can be accurately inferred as compromised or not compromised and we denote it by $P_{detect}$. 
Let us further illustrate the meaning of $P_{detect}$ using Figure~\ref{pdet}
that shows an input in reality could be either compromised or not compromised. If compromised, it can be inferred as either `compromised' with the probability $a_1$ (correct) or `undecided' with the probability $a_2$ (uncertain) or `not compromised' with the probability $a_3$ (missed detection). Similarly, if an input was not compromised, it can
be inferred as either `not compromised' with the probability $b_1$ (correct) or `undecided' with the probability $b_2$ or `compromised with the probability $b_3$ (false alarm).
Thus, for the two real cases, detection occurs with
probabilities $a_1$ and $b_1$. If an input has equal chances of
being compromised and not compromised,
then $P_{detect} =\frac{a_1 + b_1}{2}$. Else, $a_1$ and $b_1$ will
have to be weighted with their corresponding probabilities. For all practical purposes, we consider $P_{detect}$  to be at least $0.5$, since it is impractical to have a monitoring mechanism where majority of feedback are incorrect. Similarly, $P_{uncertain}=\frac{a_2+b_2}{2}$ denotes the probability of ignorance (expressing inherent uncertainty) about IoT inputs. $P_{error}=\frac{a_3+b_3}{2}$ denotes probability of errors made by the feedback system. These probabilities are used for performance evaluation.

The above features make the problem of computing the data integrity a probabilistic concept.
Hence, we compute the trust scores (utility values) as an incremental process based on observations over time slots.
If the adversary uses the same attack strategy, then the utility value will converge sooner.
On the other hand, if the adversary changes its attack strategy (i.e., dynamic attack strategy),
the utility value will oscillate even for large time windows. Later, we study a special case on how the proposed theoretical model can be modified to accommodate adversaries that do not have a fixed $P_{attack}$ and launch an On-Off attack strategy. This modified version of our method is called AWMA.

\noindent{\bf On-Off Attack:} The On-Off attack strategy is defined with an Off:On ratio. In `Off' mode the adversary does not attack. In `On' mode the adversary manipulates a random number of inputs in each time slot. It may be noted that ratios with equal Off to On modes do not depict true inconsistency. Therefore, we use higher ratios like 2:1 and 3:1. Remember that very high Off:On ratio means that the adversary behaves honest most of the time which is not realistic. 

\subsection{Data Assumptions}
\label{sec-problem}

\begin{table}[!t]
\caption{Model parameters and notations.}
\centering
\begin{tabular}{l|l}
\hline
Symbol  & Meaning \\
\hline
$\alpha$, $\beta$, $\mu$ & Not compromised, compromised, and undecided events\\
$n_{\alpha}$ & Number of data inputs detected as `not compromised'\\
$n_{\beta}$, $n_{\mu}$ & Number of data inputs detected as `compromised' and `undecided'\\
$N$ & Total number of voting components (devices)\\
$\theta_{\alpha},\theta_{\beta},\theta_{\mu}$ & Unknown probabilities for observing events $\alpha,\beta,\mu$\\
$\bar{\theta}$ & Bayesian probability parameters of the three events\\
$X(\bar{\theta})$ & Hypothesis of a event \\
$\hat{X}(\bar{\theta})$ & Posterior hypothesis or belief \\
$D(N)$ & Random vector denoting data hyper-parameter\\
$P_{detect}$ & Probability of detection of compromised data inputs \\
$P_{attack}$ & Probability of attack (attack magnitude) \\
$R_{\alpha}$, $R_{\beta}$, $R_{\mu}$ & Posterior Bayesian believes of the three events\\
$c_{n}$, $c_{c}$, $c_{u}$ & Not compromised, compromised and undecided costs \\
$c$ & Average cost of decision-making for each component \\
$\pi$ & Profit Function \\
$\pi_{P}$ & Profit Function at the Reference Point \\
$\delta$ & Deviation from Reference Point \\
$V$ & Value Function \\
$W$ & Weighting Function \\
\hline
\end{tabular}
\label{table-notations}
\end{table}

Let the three outputs of the anomaly monitoring mechanism, viz., 
`not compromised', `compromised', and `undecided' be
denoted by $\alpha$, $\beta$, and $\mu$, respectively.
Let $n_{\alpha}$ represents the number of devices inputs that have `not' been compromised,
$n_{\beta}$ be the number of 
compromised ones, and $n_{\mu}$ be the number of devices inputs for which a decision could not be made.
Of course, $n_{\alpha} + n_{\beta} + n_{\mu} = N$. Since the values of $n_{\alpha}$, $n_{\beta}$, and $n_{\mu}$ change over time,
we represent these observations at time $t$ as $n_{\alpha}(t)$, $n_{\beta}(t)$, and $n_{\mu}(t)$.

Given that the underlying parameters of the system supplying accurate data are unknown,
we use a Bayesian inference approach to incrementally update the corresponding probability
estimate for a hypothesis that the aggregate data is correct with a certain probability.
The system is only as reliable as the individual inputs are. Therefore,
we have to calculate the posterior probabilities associated with each of the possible three feedback. 
The final data integrity score will be a function of these posterior probabilities which are
also known as belief estimates in Bayesian inference.

To begin with, a uniform belief over the three possibilities is 
assumed as there is no initial information.
As time progresses, we update the belief estimates based on the 
observed values of $\alpha$, $\beta$, and $\mu$
which increases the accuracy of the estimate of the belief associated with each outcome.

We define $\theta_{\alpha}$, $\theta_{\beta}$, and $\theta_{\mu}$ as the probabilities for an input being `not compromised, `compromised', and `undecided', respectively.
Of course, $\theta_{\alpha}+\theta_{\beta}+\theta_{\mu}=1$, since the outcomes are exhaustive and mutually exclusive.
We define $X(\bar{\theta})$ as the hypothesis described by these underlying unknown {\em Bayesian probability parameters} where
  $\bar{\theta}=\{\theta_{\alpha},\theta_{\beta},\theta_{\mu}\}$.

Let $D_{\alpha}$, $D_{\beta}$, and $D_{\mu}$ denote the random variables that represent the number of times
 the outcomes $\alpha$, $\beta$, and $\mu$ occur.
 The observation data is represented as a random observation vector
 $D(N)=\{D_{\alpha}$, $D_{\beta}$, $D_{\mu}\}$. It has a multinomial distribution also known as {\em concentration hyper-parameter} of the underlying 3-tuple probability parameters described by $\theta_{\alpha}$, $\theta_{\beta}$, and $\theta_{\mu}$. The commonly used notations are tabulated in Table~\ref{table-notations}.

\section{Bayesian Inference}
\label{sec-bayes-infer}
As mentioned in Section~\ref{sec-model}, there are $N$ independently monitored components of a system whose parameters for voting
process are unknown due to changing adversarial attack strategies  and the imperfect monitoring mechanism.
Given this, we calculate the Bayesian belief associated with `not compromised'.
Similarly, we model the Bayesian posterior belief for the other two cases as well, viz., compromised and undecided.

We use the observation counts from the sequential observations
over time to calculate the posterior Bayesian estimate of each of the parameters.
Our objective is to estimate and update the probability parameters in $X(\bar{\theta})$, viz., $\theta_{\alpha}$, $\theta_{\beta}$, and $\theta_{\mu}$
based on the observation evidence $D(N)$ and prior information on the hypothesis parameter, $\bar{\theta}$.

Since there is no initial information about $\bar{\theta}$, we consider the prior parameters of $\bar{\theta}$ to be
uniformly distributed. 
Subsequent observations decide how these parameters are updated. Our first step is to calculate the Bayesian estimate of $\bar{\theta}$.

First, we show the case of estimating the posterior belief of a `not compromised' outcome ($\theta_{\alpha}$).
Since in Bayesian inference, the assumption is that prior and posterior probability
have the same distribution, we can formally define the probability parameters as:
\begin{equation}
\begin{split}
P(X(\bar{\theta})=\alpha|\bar{\theta})=\theta_{\alpha}\raisepunct{,}\\
P(X(\bar{\theta})=\beta|\bar{\theta})=\theta_{\beta}\raisepunct{,}\\
P(X(\bar{\theta}))=\mu|\bar{\theta})=\theta_{\mu}\raisepunct{.}
\label{eq7}
\end{split}
\end{equation}
This assumption is due to the well-known fact that a Dirichlet distribution acts as a conjugate prior to multinomial distributions~\cite{Sun}. Hence, prior and posterior preserve the same form.

The observations data $D(N)$ can be treated as a multinomial distribution with
the probability parameters $\theta_{\alpha}$, $\theta_{\beta}$, and $\theta{\mu}$, where the probability mass function is given by:
\begin{eqnarray}
P(D_{\alpha}=n_{\alpha},D_{\beta}=n_{\beta},D_{\mu}=n_{\mu}|\bar{\theta}) = P(D(N)|\bar{\theta}) 
= \frac{N!}{n_{\alpha}! n_{\beta}! n_{\mu}!} \theta_{\alpha}^{n_{\alpha}}\theta_{\beta}^{n_{\beta}}\theta_{\mu}^{n_{\mu}}\raisepunct{.}
\label{eq8}
\end{eqnarray}
Thereafter, we can use Bayes theorem to calculate the posterior belief estimate
on the event of a positive outcome (not compromised) $\hat{X}(\bar{\theta})=\alpha$, given observation data $D(N)$ as:
\begin{equation}
P(\hat{X}(\bar{\theta})=\alpha|D(N))=\frac{P(\hat{X}(\bar{\theta})=\alpha,D(N))}{P(D(N))}\raisepunct{.}
\label{eq9}
\end{equation}
The denominator of the above equation is the marginal probability that can be conditioned or marginalized on all possible outcomes for $\bar{\theta}$. Since probabilities are continuous, we can write:
\begin{equation}
P(D(N))= \!\!\!\!\!\!\!\!\!\!\!  \int\limits_{\qquad D(N)(\bar{\theta})} \!\!\!\!\!\!\!\!\!\!\!\! P(D(N)|\bar{\theta})f(\bar{\theta})d(\bar{\theta})\raisepunct{.}
\label{eq10}
\end{equation}
Since there is no prior information on $\bar{\theta}$ (before any observations) in Equation~(\ref{eq10}), we can assume it to be uniformly distributed such that $f(\bar{\theta})=1$. We can also put Equation~(\ref{eq8}) in Equation~(\ref{eq10}), and get:
\begin{equation}
P(D(N))= \frac{N!}{n_{\alpha}! n_{\beta}! n_{\mu}!} \!\!\!\!\!\!\!\!\!\!\!\!\!\!\!\!\!\!\!\!\!\!\!\!\!\!\!\int\limits_{\qquad\qquad\quad D(N)(\theta_{\alpha},\theta_{\beta},\theta_{\mu})} \!\!\!\!\!\!\!\!\!\!\!\!\!\!\!\!\!\!\!\!\!\!\!\!\!\!\!\!\! \!\theta_{\alpha}^{n_{\alpha}}\theta_{\beta}^{n_{\beta}}\theta_{\mu}^{n_{\mu}}d\theta_{\alpha}d\theta_{\beta}d\theta_{\mu}\raisepunct{.}
\label{eq11}
\end{equation}
After solving Equation~(\ref{eq11}) in \ref{ap1}, the denominator of Equation~(\ref{eq9}) is obtained as:
\begin{equation}
P(D(N)) = \frac{N!}{(N+2)!}\raisepunct{.}
\label{eq14}
\end{equation}
Assuming conditional
independence between the $\hat{X}(\bar{\theta})$, $D(N)$, and $\bar{\theta}$,
we calculate the numerator of Equation~(\ref{eq9}). It is explained in \ref{ap2} that how Equation~(\ref{eq17}) is obtained:
\begin{equation}
P(\hat{X}(\bar{\theta})=\alpha,D(N))=\frac{N!(n_{\alpha}+1)}{(N+3)!}\raisepunct{.}
\label{eq17}
\end{equation}
Thus, Equation~(\ref{eq9}), can be solved with dividing Equation~(\ref{eq17}) by Equation~(\ref{eq14}),
which gives:
\begin{equation}
P(\hat{X}(\bar{\theta})=\alpha| D(N))=\frac{n_{\alpha}+1}{N+3}\raisepunct{.}
\label{eq18}
\end{equation}
Similarly, $P(\hat{X}(\bar{\theta})=\beta| D(N))=\frac{n_{\beta}+1}{N+3}$ and $P(\hat{X}(\bar{\theta})=\mu| D(N))=\frac{n_{\mu}+1}{N+3}$.
These equations are the expressions for posterior belief of `not compromised', `compromised', and `undecided'.
To simplify the notations of belief estimates of the three categories, we rewrite them as $R_{\alpha}$, $R_{\beta}$, $R_{\mu}$, respectively.
Of course, it can be verified that $R_{\alpha}+R_{\beta}+R_{\mu}=1$.


\section{Data Integrity Under Uniform Attacks}
\label{sec-reliable}
In this section, we propose two types of fusion rule for the aggregate data at the IoT hub. These two models take into account different risk tolerance of different systems. Then, according to these two fusion rules, we propound our prospect theoretic approach for measuring the overall data integrity in Section~\ref{sec-integrity-PT}. At the end, we compare our prospect theoretic approach with a similar framework that uses expected utility theory in Section~\ref{sec-comparison}. 

\subsection{Optimistic System Fusion Rule}
As was discussed, in each time slot, there are three possible outcomes for the data inputs from each of the components of the IoT network: \textit{compromised, not compromised,} and \textit{undecided}. Each of these outcomes cause some costs. As it is evident, the compromised data inputs cause the highest cost, denoted by $c_{c}$. Not compromised data inputs also create some cost denoted by $c_{n}$. The third cost, denoted by $c_{u}$, is associated with the data inputs that remained undecided. Based on the system requirements, we will take different measures for undecided data inputs. The general relation between these costs is:
\begin{eqnarray}
c_{n}< c_{u}\leq c_{c}\raisepunct{.}
\end{eqnarray}

For optimistic systems, we consider half of the undecided data inputs as compromised because we assumed that the adversary has uniformly chosen the IoT inputs to attack. In other words, there is no reason for preferential attack on a certain IoT device's input. In this case $c_{u}$ is defined as follows:
\begin{equation}
c_{u}=\dfrac{c_{c}+c_{n}}{2}\raisepunct{.}
\label{eq19}
\end{equation}
Of course, when the fraction of the undecided outcomes is high, we may not be as confident about the data integrity measurement as when we have fewer undecided outcomes.

\subsection{Conservative System Fusion Rule}
Unlike the optimistic approach that the undecided outcomes are split in an equal ratio, the conservative model treats the undecided ones as if they are more likely to be compromised. In this case, we consider two weights for the compromised and not compromised costs namely $ w_{1} $ and $ w_{2}$ in a way that $ w_{1}+w_{2}=1 $ and $ 0.5<w_{2}\leq1 $. Hence,
\begin{equation}
c_{u}=w_{1}c_{n}+w_{2}c_{c}\raisepunct{.}
\label{eq20}
\end{equation}

This conservative way of computing the undecided cost is more appropriate for mission-critical systems where the 
decisions can mostly be made based on the `not compromised' inputs. Depending on how conservative a system is, we define the weights. By increasing $ w_{2} $, the chance of assuming a compromised data input as a not compromised one reduces. If the system is highly mission-critical and there is no room for risk, we consider $ w_{2}=1 $. In this case, all undecided data inputs are considered as compromised even if there could be some that were not compromised.

\subsection{Data Integrity Using Prospect Theory (PT)}
\label{sec-integrity-PT}
Using the Bayesian posterior believes of the three possible outcomes that were obtained in Section~\ref{sec-bayes-infer}, 
we want to calculate a utility value via prospect theory (PT). This utility value is calculated as follows:
\begin{eqnarray}
Utility = \sum_{i=1}^{3} V(\delta_{i})W(P_{i}) 
= V(\delta_{\alpha})W(R_{\alpha})+ V(\delta_{\beta})W(R_{\beta}) + V(\delta_{\mu})W(R_{\mu})\raisepunct{,}
\label{Utility}
\end{eqnarray}
where, $V$ denotes value function and $W$ denotes weighting function. $\delta_{\alpha}$, $\delta_{\beta}$, and $\delta_{\mu}$ are three deviation values assigned to the three independent outcomes of collected data from each device in an IoT network. These deviations show the difference between profit 
function, denoted by $ \pi $, and reference point denoted by $\pi_{P}$. Due to the independence of the outcomes, a separate profit function is defined for each of them. These profit functions are denoted by $ \pi_{\alpha} $, $ \pi_{\beta} $, and $ \pi_{\mu} $. The deviation values and profit functions are defined as follows:
\begin{eqnarray}
\left\{ \begin{array}{ll}
\delta_{\alpha}=\pi_{\alpha}-\pi_{p},\qquad \hspace{-0.35mm} \text{where} \quad \pi_{\alpha}=N c - n_{\alpha}c_{n}; \\
\delta_{\beta}=\pi_{\beta}-\pi_{p},\qquad \text{where} \quad \pi_{\beta}=N c - n_{\beta}c_{c};\\
\delta_{\mu}=\pi_{\mu}-\pi_{p},\qquad \text{where} \quad \pi_{\mu}= N c - n_{\mu}c_{u}.\\
\end{array} \right.
\label{myeq5}
\end{eqnarray}
In order to to calculate deviation values, We also need to define the reference point, denoted by $\pi_{P}$. It is initialized by assuming all the IoT data inputs as \textit{not compromised}:
\begin{eqnarray}
\pi_{p}=N(c-c_{n})\raisepunct{.}
\label{myeq3}
\end{eqnarray}

According to Equations~(\ref{myeq5}) and (\ref{myeq3}), the state (hypothetical outcome) which is considered for data inputs of all $ N $ devices at the reference point determines whether the real outcomes are gains or losses. By considering the data inputs of all $N$ devices as \textit{not compromised} at the reference point, any data input which is \textit{not compromised} will always be considered as a gain and other outcomes will be counted as a gain or loss based on the cost values ($c_{c}$ and $c_{u}$). The reason for always considering \textit{not compromised} outcomes as gain is that the \textit{not compromised} data inputs have the lowest cost ($c_{n}$). However, if we had considered all the data inputs as \textit{compromised} or \textit{undecided} at the reference point, the value of $ \delta $ for each outcome would increase such that even compromised data inputs yield a gain.
 
\begin{figure} [h]
\begin{center}
\includegraphics[width=3.5in, height=2in, trim=0.2cm 0 0 0cm]
{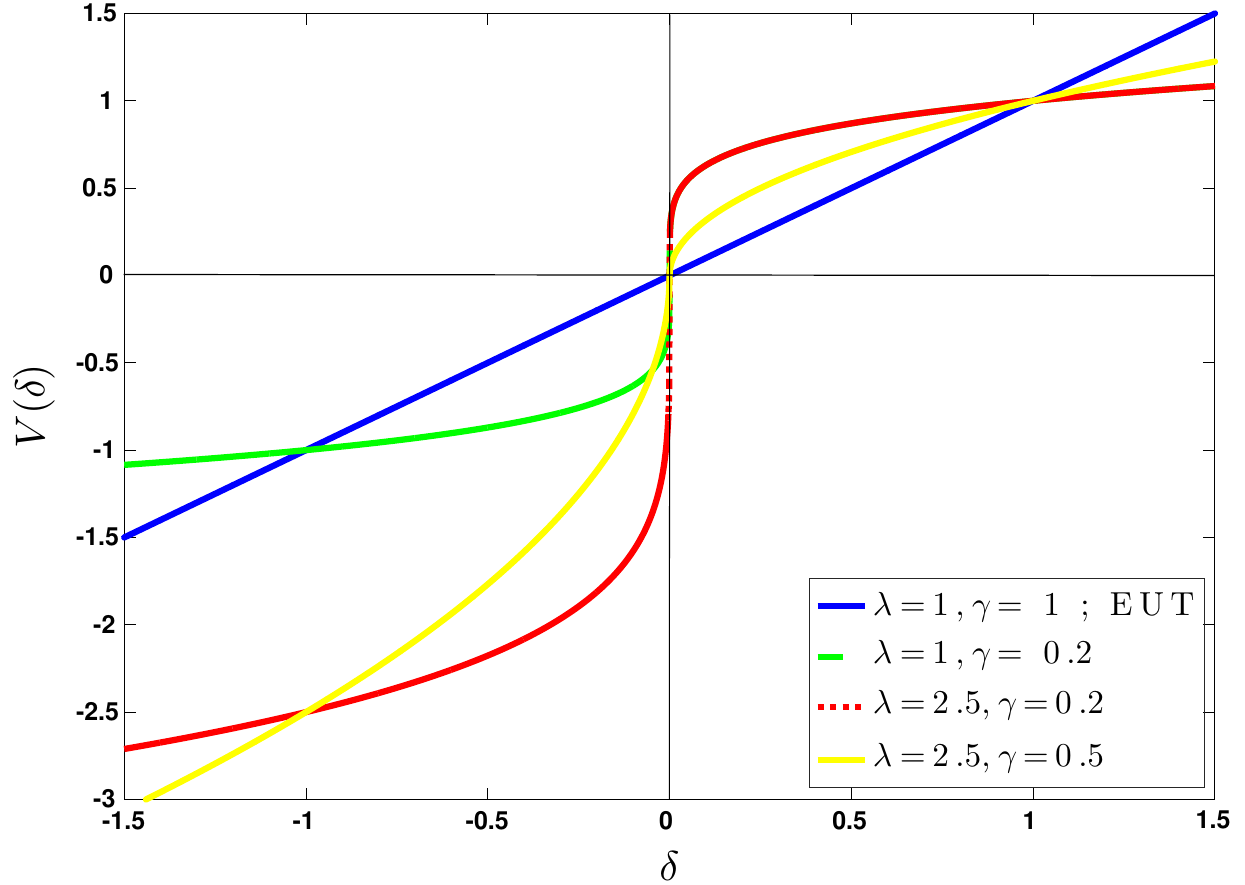}
\caption{Examples of value function with different $\lambda$ and $\gamma$.}
\label{value function}
\end{center}
\end{figure}

Value function ($V(\delta)$), shown in Figure~\ref{value function}, is an asymmetrical S-shaped function. It is asymmetric because of its loss aversion nature which causes the same absolute deviation values ($\delta$) to have more impact on the loss than on the gain. Its value is dependent on the deviation of the profit values from the reference point, defined in Equation~(\ref{myeq5}). Value function is obtained as follows: 
\begin{eqnarray}
   V(\delta) =\left\{ \begin{array}{ll}
	 \delta^{\gamma}, & \mbox{if $\quad \delta \geq 0 \,$};\\
	
	 -\lambda(-\delta)^{\gamma}, & \mbox{if $\quad \delta < 0$}\raisepunct{.}\end{array} \right.
\label{myeq4}
\end{eqnarray}

Positive part of the value function represents the gain and its negative section denotes the loss in reliability of the collected data from the IoT devices. $\lambda$ and $\gamma$ are two parameters used for  controlling loss aversion and risk aversion where $\lambda > 1$, and $0 < \gamma <1$ ~\cite{mehrdad}. By increasing $\lambda$, the IoT network will become more loss averse and consequently will become more asymmetric with the loss part becoming more convex. By decreasing parameter $\gamma$, the IoT network will become more risk averse. The effect of these parameters on value function is shown in Figure~\ref{value function}. Choosing the right values for these two parameters depends on the IoT network. As the network becomes more conservative, it becomes more loss averse.

According to Equation~(\ref{Utility}), we also need to define the weighting function to calculate the utility value. Based on  prospect theory, in real life decision-making process, people overreact to lower probabilities and under-react to higher probabilities~\cite{prospect}. Here, we have the same situation. For example, if the data input from one device is compromised, it will have a significant impact on reliability of the aggregate data. However, if we have $30$ devices and data inputs of $20$ of them are compromised, having one more compromised data input will not cause a significant difference. Furthermore, the effect of probability weights is not the same for loss and gain~\cite{prospectweight}. As it is obvious, it is desirable to have the minimum possible loss rather than achieving a huge gain since even a little loss will result in  losing confidence in the aggregate data. Therefore, two similar weighting functions but with two different parameters, $ \omega $ and $ \varphi $, are defined for gain denoted by $W^{+}(p)$ and loss denoted by $W^{-}(p)$ as follows:
\begin{eqnarray}
\left\{ \begin{array}{ll}
W^{+}(p)=\frac{p^{\varphi}}{[p^{\varphi}+(1-p)^{\varphi}]^{\frac{1}{\varphi}}}\,\raisepunct{,}\hspace{1mm}\quad \;\;\, \text{where} \quad 0.5\leq\varphi \hspace{0.5mm}<1;  \\
W^{-}(p)=\frac{p^{\omega}}{[p^{\omega}+(1-p)^{\omega}]^{\frac{1}{\omega}}}\raisepunct{,}\qquad \text{where} \quad 0.5\leq\omega <1\raisepunct{,}
\end{array} \right.
\label{myeq6}
\end{eqnarray}
where, $ p $ corresponds to the posterior belief of the three possible outcomes

\begin{figure} [h]
\begin{center}
\includegraphics[width=3.5in, height=2in,trim=0.15cm 0 0 0cm]{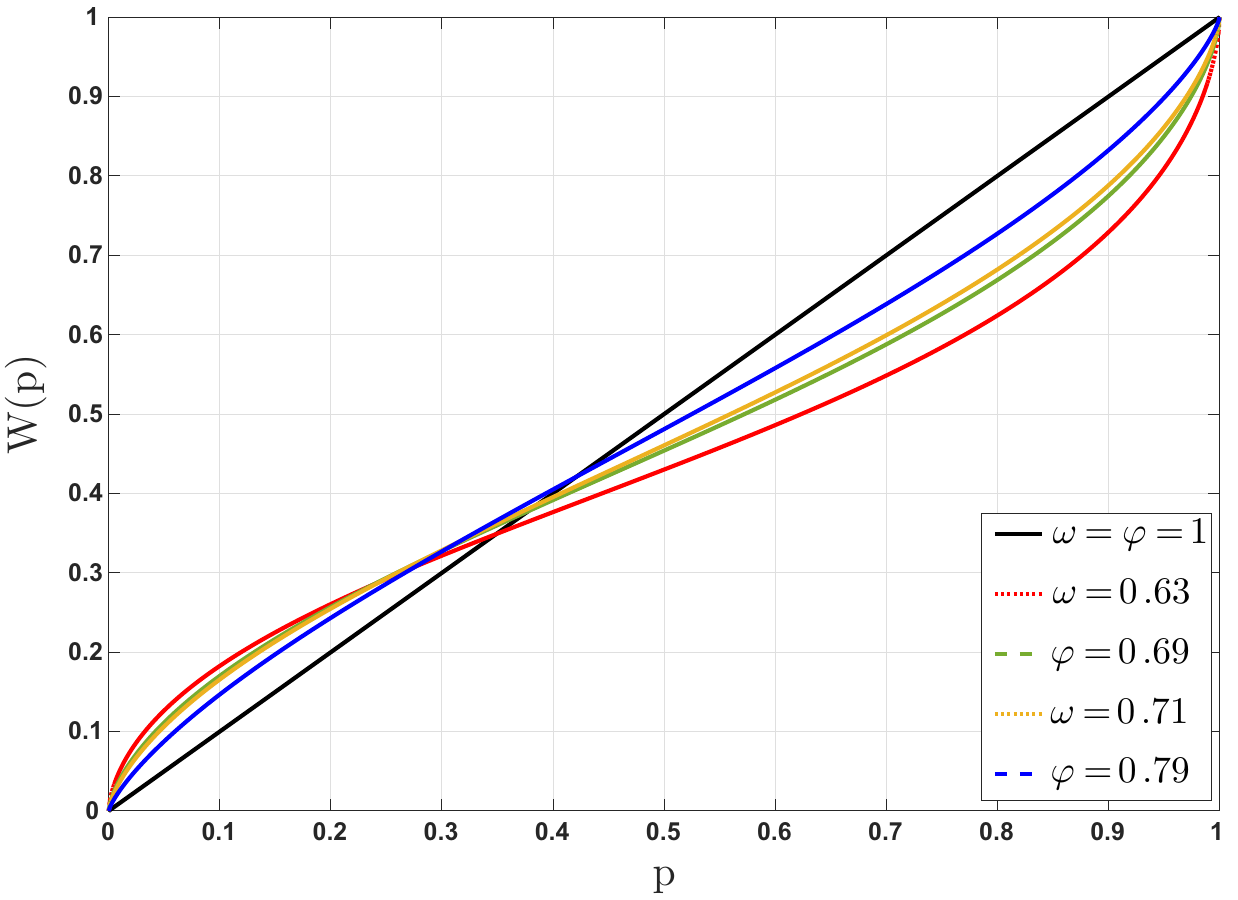}
\caption{Effect of $\varphi$ and $\omega$ on weighting function.}
\label{weight function}
\end{center}
\end{figure}

Parameters $\varphi$ and $\omega$ are defined in a way to emphasize the loss which means choosing lower values for $ \omega $ in comparison with $ \varphi $. Their values also depend on the system that we are dealing with. In conservative systems, the effect of weighting function for loss will be higher than optimistic systems. Therefore, conservative systems have lower values of $\omega$ in order to cause strict penalty for loss. On the other hand, the value of $\varphi$ for conservative systems is larger than the optimistic ones since achieving gain in conservative systems is not as highly valued as in optimistic systems. In other words, achieving gain in a conservative system will not achieve a positive utility as high as its optimistic counterpart. The effects of these two parameters are demonstrated in Figure~\ref{weight function}.

By using value function ($V(\delta)$) and weighting function ($W(p)$), we obtain a utility value for an IoT network in each time slot. This utility value indicates the reliability of the aggregate data in that IoT network. It should be noticed that each IoT network has its own parameters according to its degree of risk tolerance for calculating the utility value. In addition, a threshold can be defined to determine a limit for the utility value of an IoT network to be acceptable. If the calculated utility is higher than this threshold, the aggregate data in that network has acceptable integrity.

\subsection{Data Integrity Using Expected Utility Theory (EUT)}
\label{sec-comparison}
Though we proposed a prospect theoretic approach for evaluating data integrity in IoT networks, there are other
competing utility theories and one of the most popular ones is the expected utility theory (EUT)~\cite{mongin1997expected}. Expected utility theory can be imagined as a special case of prospect theory (PT) where we have no weights for weighting function, $\varphi = \omega$, and also $\lambda = \gamma$ in value function.
The utility derived from EUT is not as risk averse and loss averse as that obtained by prospect theory~\cite{schechter2007risk}. Therefore, it is expected that it cannot significantly differentiate between optimistic and conservative systems.

When we compare PT with EUT, we should analyze two functions: value function and weighting function, which are shown accordingly in Figures~\ref{value function} and~\ref{weight function}. In addition, we are dealing with three possible outcomes in this problem. According to Equation~\ref{Utility}, the final utility is the summation of the utilities of these three outcomes. Therefore, we should separately analyze the effect of the two mentioned functions on each of these three utilities. These effects depend on the adversary's attack magnitude.


\subsubsection{Small Attack Magnitudes:}
By taking into account Figure~\ref{value function} and Equation~\ref{myeq5}, for small attack magnitudes, it is expected that compromised and undecided data inputs will encounter higher negative values for value function of PT than EUT. In addition, since these data inputs have small $ p $, they will have larger weights for PT than EUT according to Figure~\ref{weight function}. On the contrary, not compromised data inputs will have larger positive values for value function in PT than EUT. On the other hand, when the attack magnitude is small, the not compromised data inputs will have a higher probability ($p$ in Figure~\ref{weight function}). Therefore, according to Figure~\ref{weight function}, their weights for PT is less than EUT. By considering all of these facts about small attack magnitudes, it is expected that the total utility value, which is a positive value, is larger for EUT. 

\subsubsection{Moderate Attack Magnitudes:}
However, for moderate attack magnitudes, the weights for PT and EUT are almost the same and the only difference is in the value function. According to Figure~\ref{value function}, the positive value function for PT and EUT are not significantly different. However, the negative part is noticeably different. Therefore, the positive utility caused by not compromised data inputs is almost the same for PT and EUT but the absolute value of negative part of PT which is caused by compromised and undecided data inputs is more than EUT. In conclusion, for moderate attack magnitudes, it is expected that the absolute utility value of PT is larger than EUT. 

\subsubsection{Large Attack Magnitudes:}
For large attack magnitudes, similar to middle attack magnitudes, the positive part of value function is almost the same for PT and EUT. However, for the negative part which is caused mostly by compromised data inputs, the story is different. As mentioned earlier, PT is risk seeking in the loss domain. It means that for large negative values of $x$ in value function, its absolute value for EUT is larger than PT. Therefore, the magnitude of negative value function caused by compromised data inputs in EUT is more than PT. In addition, the negative values of value function caused by compromised data inputs have larger weights in EUT. Therefore, it is expected that EUT will have higher absolute value than PT for large attack magnitudes. 

\subsubsection{Conservative Systems vs Optimistic Systems:}
Neglecting the under-reaction to larger probabilities in expected utility theory is more noticeable for conservative systems which have higher cost for malicious nodes. It means that since the expected utility does not utilize a weighting function, the effect of the probabilities for large probabilities in  EUT is more than PT. Thus, the absolute utility value of EUT is larger than PT. The effect of this phenomenon is to the extent that for high attack magnitudes, utility value of EUT for optimistic systems is lower than the utility value of PT for conservative systems.

\section{Data Integrity Under On-Off Attacks}
A reliable trust management mechanism should be immune to On-Off attacks. Therefore, in this section, we propose an extension of our method called Asymmetric Weighted Moving Average (AWMA). This model is adaptive to On-Off attacks for measuring the trust score.

\subsection{Existing Trust Update Schemes: CWMA and EWMA}
\label{on-off-model}
In On-Off attacks, the adversaries have preferences over time periods where they may choose not to attack for some time (Off mode) and then attack for some time with a random magnitude (On mode). In such a case, either Cumulative Weighted Moving Average (CWMA) or Exponentially Weighted Moving Average (EWMA) would not reflect true behavior of the aggregate data. CWMA will lag in reflecting such attacks, and EWMA will allow the system to quickly recover or redeem its reputation when the adversary switches back to honest behavior. 

\begin{figure} [h]
\begin{center}
\includegraphics[width=3.5in, height=2.3in]{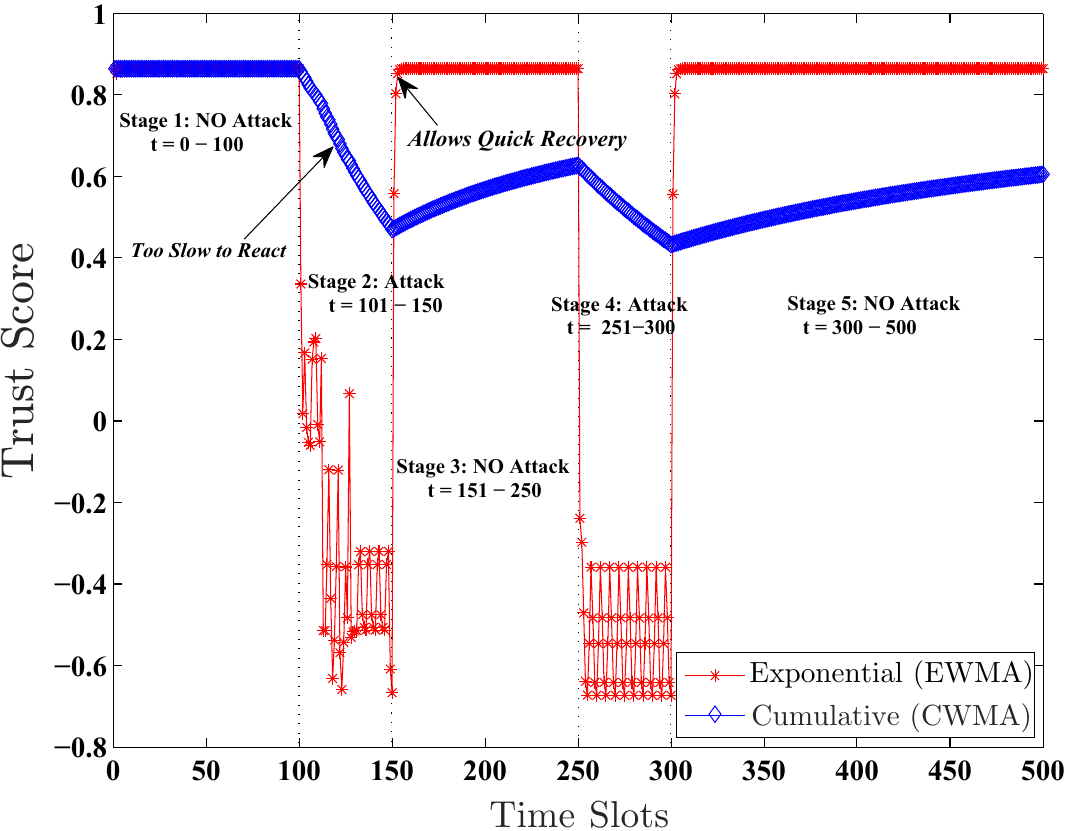}
\caption{Shortcomings of CWMA and EWMA.}
\label{problems_with_existing}
\end{center}
\end{figure}

Under On-Off attacks, the data integrity scoring mechanism should not allow the system to recover its trust score even though the adversary starts behaving well after a short burst of attack. In Figure~\ref{problems_with_existing}, we show an example where the adversary employs a 2:1 Off-On ratio where it divides the time domain of $500$ slots to five stages; `Stage 1' in time range of $t=0-100$, `Stage 2' in time range of $t=101-150$, `Stage 3' in the time range of $t=151-250$, `Stage 4' in time range of $t=251-300$, and `Stage 5' in time range of $t=301-500$. stage 5 is a no attack phase in order to analyze the after effects of On-Off attacks. 

In Stage 1, the adversary does not attack in a bid to gain high trust of the system. In Stage 2, it attacks for the next 50 time slots with a random magnitude in each of the time slots. In Stage 3, it does not attack in any of the $100$ time slots. In Stage 4, it again attacks in $50$ time slots with a random attack magnitude. In Stage 5, it again behaves cooperatively. 

Suppose an algorithm that checks whether a system is compromised or not every 50 time slots and uses the threshold of zero. If the trust score goes below this threshold, the aggregate data should be considered unusable for decision-making. in the time slots $100^{th}$, $200^{th}$, $250^{th}$, $350^{th}$, $400^{th}$, $450^{th}$ and $500^{th}$, the system's data is deemed usable by both trust scoring schemes (CWMA and EWMA) although the adversary is employing a stealthy On-Off attack. We see that CWMA reacts too slowly and fails to reflect malicious activity even at the end of the Stage 2. On the other hand, EWMA detects attacks quickly but also allows such nodes to quickly recover their reputation in $151^{th}$ and $301^{th}$ time slots in the ensuing Off period. Hence, there is a need for a special trust update scheme that would restrict the average data integrity from improving quickly when the adversary starts behaving cooperatively after  malicious activity. In addition, it should be responsive enough to decrease the integrity score as soon as the adversary starts acting maliciously after building a high reputation.

\subsection{Weighted Trust Score}
	
As was discussed, in each time slot, we calculate a utility value for the aggregate data using Equation~\ref{Utility}. Let us denote the utility value at time $ t $ by $u_{di}(t)$ which indicates the data integrity in that time slot. However, this utility value has an infinite range, ($-\infty,\infty$), which is not applicable to On-Off attacks. In trust scoring for On-Off attacks, we need a trust management scheme that can quickly respond to attacks (in {\em On} mode) and gradually recover the trust score when the node shows a good behavior for a while (in {\em Off} mode). Thus, for decision-making about current time slot, we need to consider the node's behavior in previous time slots. However, it is not possible to capture the effect of data integrity during previous time slots in current time slot if they do not have normalized weights. Therefore, we need to map the utility values to a bounded range. So, we define weighted trust score denoted by $w_{di}(t)$ using Equation~(\ref{final_rating}). In each time slot $t$, this function maps utility values obtained by Equation~\ref{Utility} to the bounded range of $[-1,1]$.
\begin{eqnarray}
w_{di}(t) =\left\{ \begin{array}{ll}
	 1-e^{-|u_{di}(t)|},\;\; & \mbox{if $\quad u_{di}(t) > 0 $};\\

	 -(1-e^{-|u_{di}(t)|}),\;\; & \mbox{if $\quad u_{di}(t) < 0$};\\
	 0,\;\; & \mbox{if $\quad u_{di}(t) = 0$}.\end{array} \right.
\label{final_rating}
\end{eqnarray}
It should be noted that the weighted trust scores monotonically increase with increase of utility values.


	
\subsection{Asymmetric Weighted Moving Average (AWMA)}
\label{sec-AWMA}

In order to update trust scores in On-Off attacks, we should take into account both the behavior of nodes during the current time slot and the historical behavior of nodes during previous time slots. In this regard, we propose Asymmetric Weighted Moving Average (AWMA) scheme using weighted trust scores obtained from Equation~\ref{final_rating}. This scheme uses both the weighted trust score during the current time slot ($w_{di}(t)$) and the cumulative moving average of trust scores before that time slot ($w_{di}^{mavg}(t-1)$) to update the current cumulative moving average ($w_{di}^{mavg}(t)$). $w_{di}^{mavg}(t)$ indicates the data integrity of IoT network at time $ t $. However, AWMA does not use traditional cumulative moving average. AWMA is based on the socially inspired concept that bad actions are far more remembered than good actions. This fact establishes the basis of AWMA scheme where current weighted trust score and cumulative moving average trust score are given different weights under different scenarios with respect to a threshold, denoted by $\Gamma_{on-off}$.
The value of $\,\Gamma_{on-off}$ is dictated by the system specific risk attitude which defines what can be considered as sufficiently good behavior. Different scenarios are differentiated by defining four weighting factors $\chi_{a}$, $\chi_{b_{max}}$, $\chi_{c_{min}}$ and
 $\chi_{d}$ such that $0< \chi_{a} < 1$, $0 << \chi_{b_{max}} < 1$, $0 < \chi_{c_{min}} << 1$, and $0 < \chi_{d} < 1$. 
It should be noticed that since $\chi_{c_{min}}$ is significantly smaller than $\chi_{b_{max}}$, it causes an asymmetry.\vspace{2mm} Now, there are four possible scenarios at time $t$ with regard to the threshold:
\vspace{2mm}
Case (a): $\;\;w_{di}^{mavg}(t-1) > \Gamma_{on-off}\,,\quad$ and $\quad\; w_{di}(t) > \Gamma_{on-off}\;$; \\
\vspace{2mm}
Case (b): $\;\;w_{di}^{mavg}(t-1) > \Gamma_{on-off}\,,\quad$ and $\quad\; w_{di}(t) \leq \Gamma_{on-off}\;$; \\
\vspace{2mm}
Case (c): $\;\;w_{di}^{mavg}(t-1) \leq \Gamma_{on-off}\,,\quad$ and $\quad\; w_{di}(t) > \Gamma_{on-off}\;$; \\
\vspace{2mm}
Case (d): $\;\;w_{di}^{mavg}(t-1) \leq \Gamma_{on-off}\,,\quad$ and $\quad\; w_{di}(t) \leq \Gamma_{on-off}\;$.

In Case (a), a cumulative moving average higher than $\Gamma_{on-off}$ suggests that the system is maintaining a sufficiently good behavior. If the current weighted trust score is also higher than $\Gamma_{on-off}$, it suggests continuity of the good behavior. Hence, continuing good behavior is rewarded with a high weighting factor ($\chi_{a}$) to $w_{di}(t)$  and a low weighting factor ($1-\chi_{a}$) to $w_{di}^{mavg}(t-1)$. We name $\chi_{a}$ as the {\em rewarding factor} such that $1 > \chi_{a} > 0 $. It helps a historically reliable system to improve, or at least maintain its reputation if it also behaves in a cooperative manner in time slot $t$. Thus, for Case (a), cumulative moving average is updated as:
\begin{eqnarray}
  w_{di}^{mavg}(t) = (1-\chi_{a})\times w_{di}^{mavg}(t-1) + \chi_{a}\times w_{di}(t)\raisepunct{.} 
\end{eqnarray}

In Case (b), a cumulative moving average higher than $\Gamma_{on-off}$ and a current weighted trust score less than $\Gamma_{on-off}$ suggests that the system is maintaining a sufficiently good behavior up to time $t-1$ and then initiated some anomalous behavior. Hence, all the good behaviors until now need to be forgotten and a very high weight be given to the
current slot's anomalous behavior. This will force the system's cumulative moving average to quickly decrease. Once this happens, Case (c) would ensure that the cumulative moving average is not able to redeem itself quickly. Therefore, $w_{di}(t)$ is assigned a large weight ($\chi_{b_{max}}$) such that $1 > \chi_{b_{max}} >> 0 $ and $w_{di}^{mavg(t-1)}$ is weighted with $1-\chi_{b_{max}}$. We name $\chi_{b_{max}}$ as the {\em punishment factor}. The larger the value of punishment factor, the quicker the drop of the trust score. Hence, the system's reaction to new evidence of malicious behavior will be more severe. In such cases,
the cumulative moving average is updated as:
\begin{eqnarray}
 w_{di}^{mavg}(t) = (1-\chi_{b_{max}})\times w_{di}^{mavg}(t-1) + \chi_{b_{max}}\times w_{di}(t)\raisepunct{.} 
\end{eqnarray}

In Case (c), a cumulative moving average lower than $\Gamma_{on-off}$ but a current weighted trust score higher than $\Gamma_{on-off}$ signifies a system where current inputs are cooperative but has a history of anomalous behavior which may be as recent as $t-1$. Therefore, we assign $w_{di}(t)$ a very small weight $\chi_{c_{min}}$ such that $0 < \chi_{c_{min}} << 1$ to cancel out its large value. In addition, we assign a large weight $1-\chi_{c_{min}}$ to emphasize $w_{di}^{mavg(t-1)}$. We name $\chi_{c_{min}}$ as the {\em redemption factor} that controls how fast or slow a system with malicious history can redeem its trustworthiness if it shows good behavior for a sufficiently long time. Redemption factor also makes it possible for systems in noisy environments to redeem their trust scores. A low redemption factor ensures that the trust score is not increased quickly even though a system starts to behave honestly after a period of malicious behavior. In this case, cumulative moving average is updated as:
\begin{equation}
 w_{di}^{mavg}(t) = (1-\chi_{c_{min}})\times w_{di}^{mavg}(t-1) + \chi_{c_{min}}\times w_{di}(t)\raisepunct{.} 
\end{equation}

In Case (d), both cumulative moving average and current weighted trust score are below $\Gamma_{on-off}$. It indicates anomalous behavior in both current time slot and previous time slots. In such a case, we define $\chi_{d}$ as the {\em retrogression factor} which is assigned as the weight to the current weighted trust score. Also, $1-\chi_{d}$ is assigned to cumulative moving average. So, the cumulative moving average is updated as:
\begin{equation}
    w_{di}^{mavg}(t) = (1-\chi_{d})\times w_{di}^{mavg}(t-1) + \chi_{d}\times w_{di}(t)\raisepunct{.} 
\end{equation}

Unlike cumulative and exponential weighted moving averages, our proposed asymmetric weighted moving average (AWMA) scheme is effective for trust management during On-Off attacks. In Section~\ref{sec-trustmanage}, we illustrate how AWMA scheme works by comparing it with CWMA and EWMA. Also, we discuss that AWMA is effective in distinguishing malicious IoT data inputs for networks experiencing intermittent noise.

\section{Model Simulation and Results}
\label{sec-sim}
We simulate a generic IoT network with $100$ IoT devices.
Inputs from all devices are monitored by an imperfect monitoring mechanism
that produces three possible outcomes in each time slot. We study different probability of detection and probability of attack to capture their effects on data integrity measurements. We plot instantaneous and moving average of utility values (trust scores). For all the simulations, the prospect theoretic parameters are fixed as follows: $\lambda=2$, $\gamma=0.5$, $\omega=0.63$, and $\varphi=0.69$.

Under uniform (non-opportunistic) attacks, an adversary attacks different sets of data inputs over time. In addition, the number of compromised inputs may vary over each time slot. However, the long-term average of number of compromised inputs, denoted by $P_{attack}$, remains the same. On the other hand, under On-Off (opportunistic) attacks, $P_{attack}$ becomes meaningless since the adversary behavior is not monotonous. Therefore, in simulations, we need to analyze how our proposed AWMA method reacts to different modes of attack (On or Off) for measuring the trustworthiness of the aggregate data.

\begin{figure} [!t]
\begin{center}
\includegraphics[width=3.5in, height=2in,trim=0.15cm 0 0 0cm]{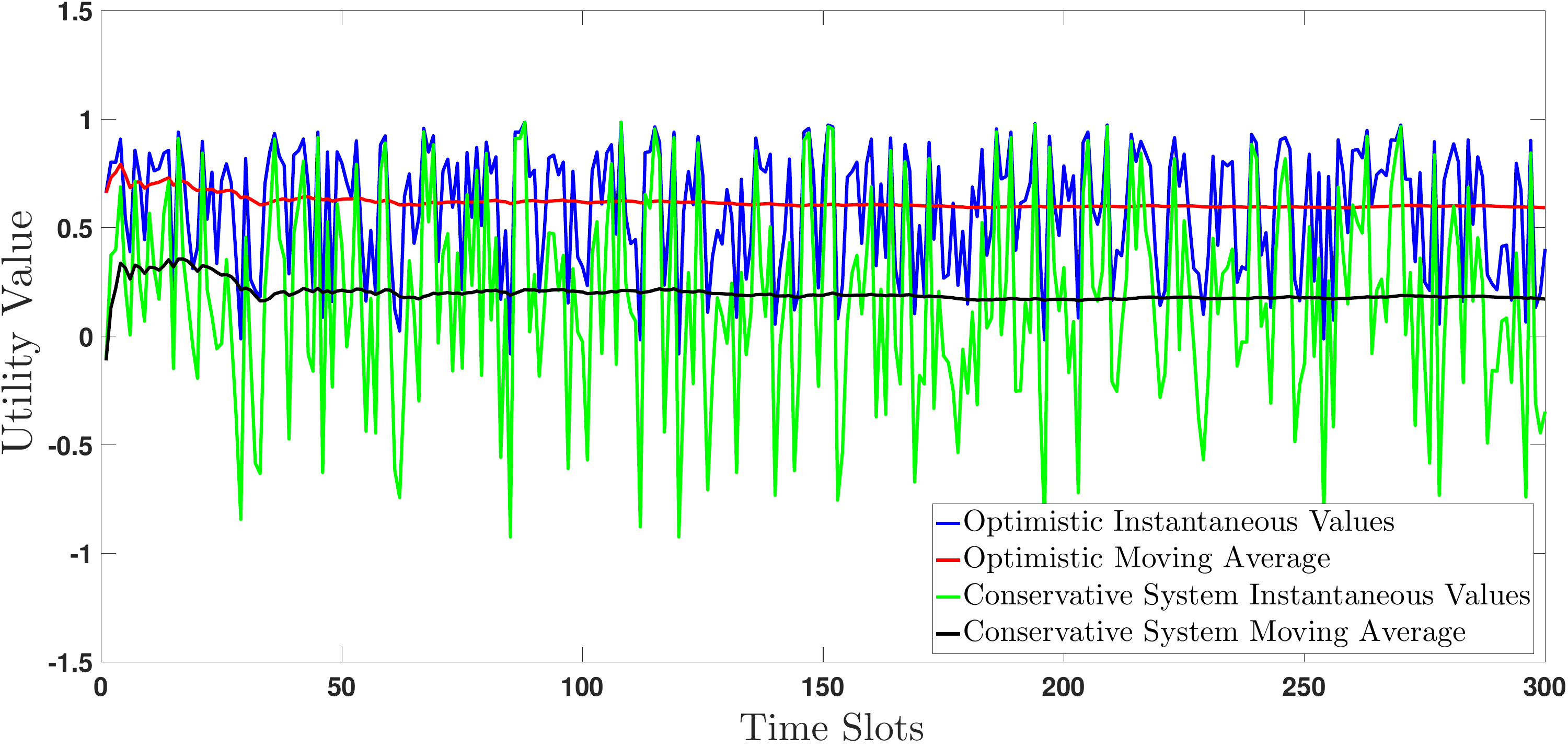}
\caption{Instantaneous and Moving Average utility values of optimistic and conservative systems over time under same attack and detection conditions: $P_{attack}=0.1$ and $P_{detect}=0.9$.}
\label{reliability_over_time}
\end{center}
\end{figure}

\begin{figure} [!b]
\begin{center}
\includegraphics[width=4in, height=2.3in,trim=0.15cm 0 0 0cm]{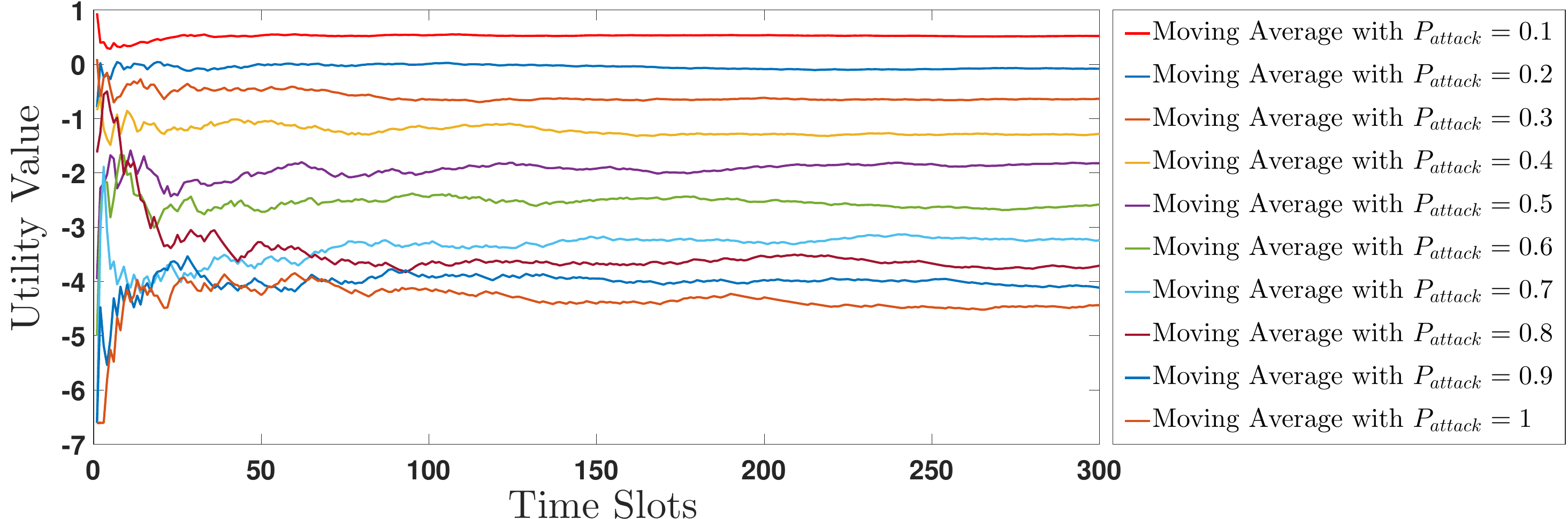}
\caption{Average utility values of an optimistic system over time under different attack magnitudes.}
\label{optimistic_pa}
\end{center}
\end{figure}

\subsection{Optimistic and Conservative Systems Under Same Attack Conditions}
In Figure~\ref{reliability_over_time}, we plot the instantaneous and steady state utility values (trust scores) for
both optimistic and conservative systems when the adversary launches a uniform attack with $P_{attack}=0.1$ and the system is able to detect compromised data inputs with $P_{detect}=0.9$. The optimistic system is designed with following costs: $c_{c}=0.1$, $c_{n}=0.01$, and $c_{u}=0.055$. The costs for conservative system are defined as follows: $c_{c}=0.1$, $c_{n}=0.01$, and $c_{u}=0.09$.
 
We observe that the instantaneous utility values fluctuate over time due to the particular realizations of $P_{attack}$ and imperfect anomaly monitoring ($P_{detect}$) in each time slot.
With sufficient observations, the moving average of optimistic utility values converges to a positive steady state value around $0.5$. Furthermore, as expected, the moving average of conservative system utility values converges to a steady state value lower than the corresponding value related to optimistic system. This value is almost zero. Therefore, under the same attack condition ($P_{attack}$) and anomaly monitoring mechanism ($P_{detect}$), the aggregate data from an optimistic system is usually reliable but the aggregate data from a conservative system is mostly unreliable.

\subsection{Utility Value and Attack Magnitude}
\begin{figure}[!t]
\begin{center}
\subfigure[\label{high_pdet}]{
\epsfig{figure=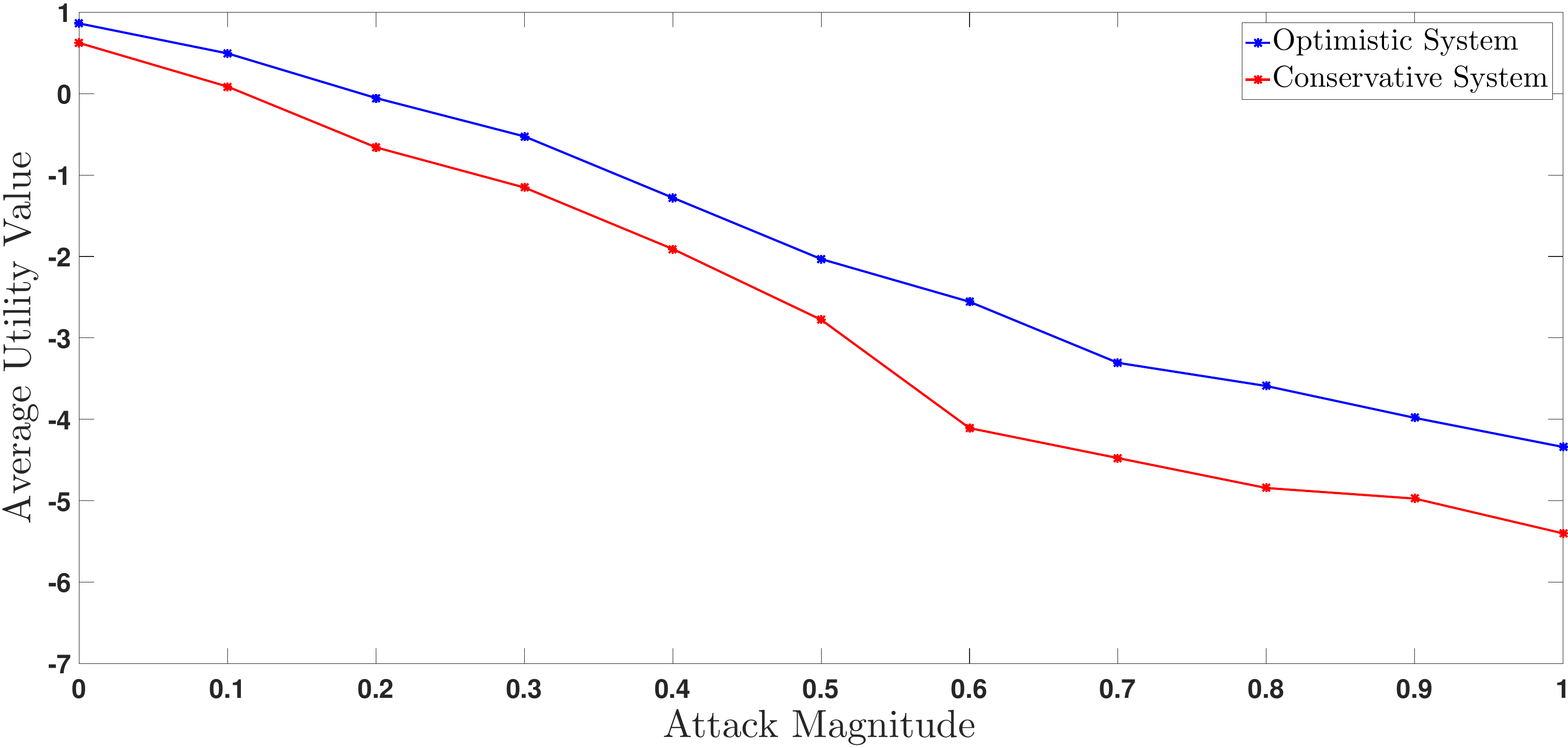,width=3.5in,height=2in}}
\subfigure[\label{low_pdet}]{
\epsfig{figure=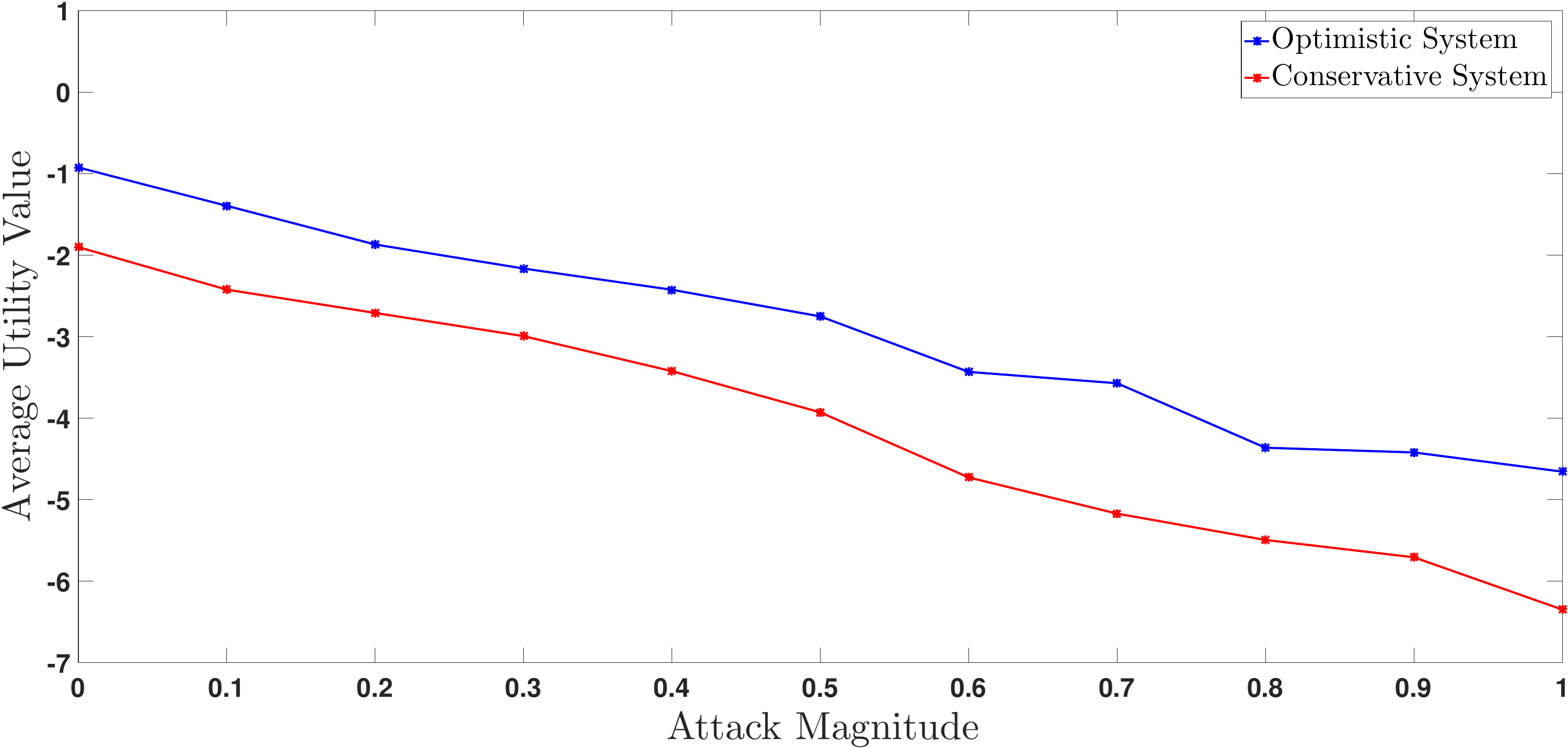,width=3.5in,height=2in}}
\caption{Average utility values of optimistic and conservative systems over time vs attack magnitude under two different $P_{detect}$. (a) $P_{detect}$ = 0.9 , (b) $P_{detect}$ = 0.5.}
\label{op_vs_con}
\end{center}
\end{figure}

In Figure~\ref{optimistic_pa}, we plot the steady state utility values in an optimistic system for different attack magnitudes (from $P_{attack}=0.1$ to $P_{attack}=1$) under the same monitoring mechanism ($P_{detect}=0.9$). In Figure~\ref{op_vs_con}, we illustrate the average utility values for both optimistic and conservative systems under different attack magnitudes ($P_{attack}$) using two different monitoring mechanisms $P_{detect}$.

According to Figure~\ref{optimistic_pa} and Figure~\ref{op_vs_con}, aggregate data in an optimistic system with a low $P_{attack}$ is reliable most of the time. However, if the attack magnitude increases, the aggregate data becomes mostly unreliable. On the other hand, the aggregate data in a conservative system is mostly unreliable even with small attack magnitudes. Nevertheless,in a conservative system, with an increase in attack magnitude, it becomes more unreliable.

In addition, Figure~\ref{op_vs_con} reveals that with a reduction in $P_{detect}$, the effect of attack magnitude on utility values follows the same trend with just a small decrease in utility values. The effect of attack magnitude on utility values is almost linear. Furthermore, as $P_{detect}$ decreases, the difference between average utility values of an optimistic system and a conservative system increases because of higher cost for undecided data inputs ($c_u$) in conservative systems.


\subsection{Utility Value and Imperfect Monitoring}
\begin{figure} [!t]
\begin{center}
\includegraphics[width=3.5in, height=2.5in,trim=0.15cm 0 0 0cm]{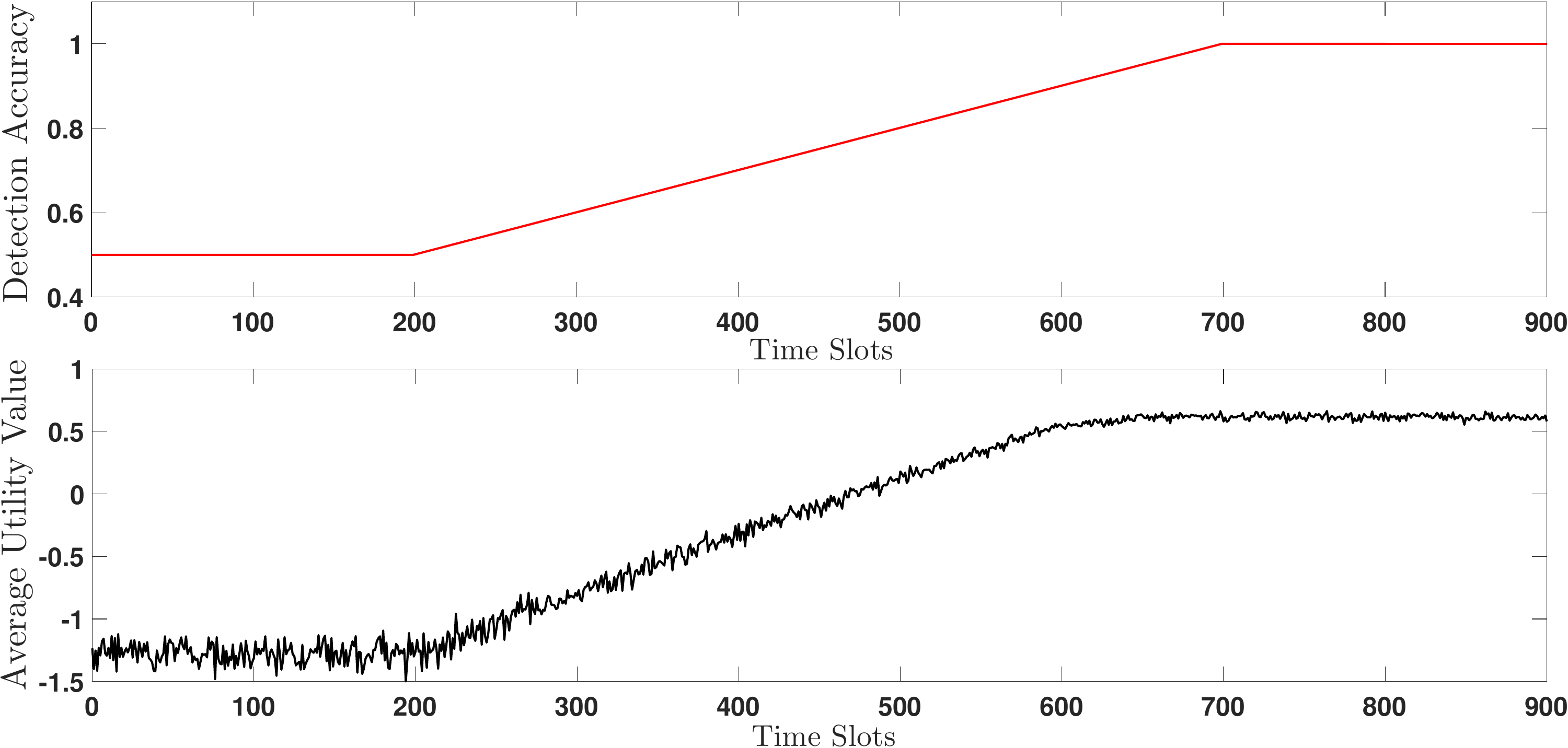}
\caption{Average utility values of an optimistic system over time with increasing detection accuracy ($P_{detect}$).}
\label{different_pd}
\end{center}
\end{figure}

Imperfect monitoring can have a negative effect on utility values and subsequently on reliability of aggregate data as much as the attack magnitude. Figure~\ref{different_pd} shows the impact of imperfect monitoring on an optimistic system when the attack magnitude remains the same ($P_{attack}=0.1$) and $P_{detect}$ increases over time. We observe that by an increase in $P_{detect}$, the aggregate data becomes more reliable with a linear trend.


\subsection{Prospect Theory VS Expected Utility Theory}
\label{sec-ptVSeut-simulation}
\begin{figure} [h]
\begin{center}
\includegraphics[width=5in, height=2in]{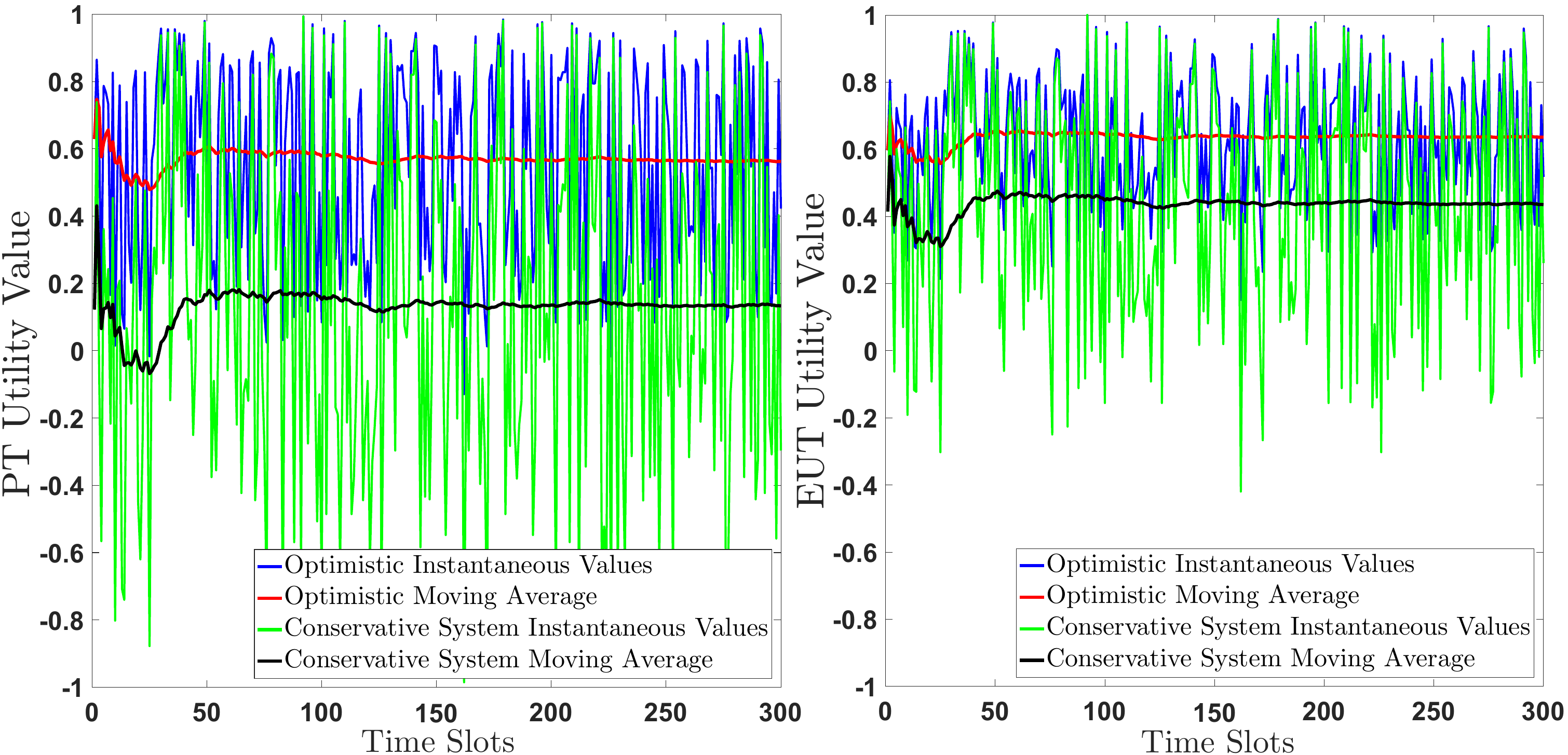}
\caption{Instantaneous and Moving Average utility values of optimistic and conservative systems over time obtained by prospect theory and expected utility theory under same attack and detection conditions: $P_{attack}=0.1$ and $P_{detect}=0.9$.}
\label{EUT1}
\end{center}
\end{figure}

Expected utility is not as risk averse and loss averse as prospect theory. Therefore, as shown in Figure~\ref{EUT1}, it does not significantly differentiate between optimistic and conservative systems.

\begin{figure} [!b]
\begin{center}
\includegraphics[width=3.5in, height=2in,trim=0.15cm 0 0 0cm]{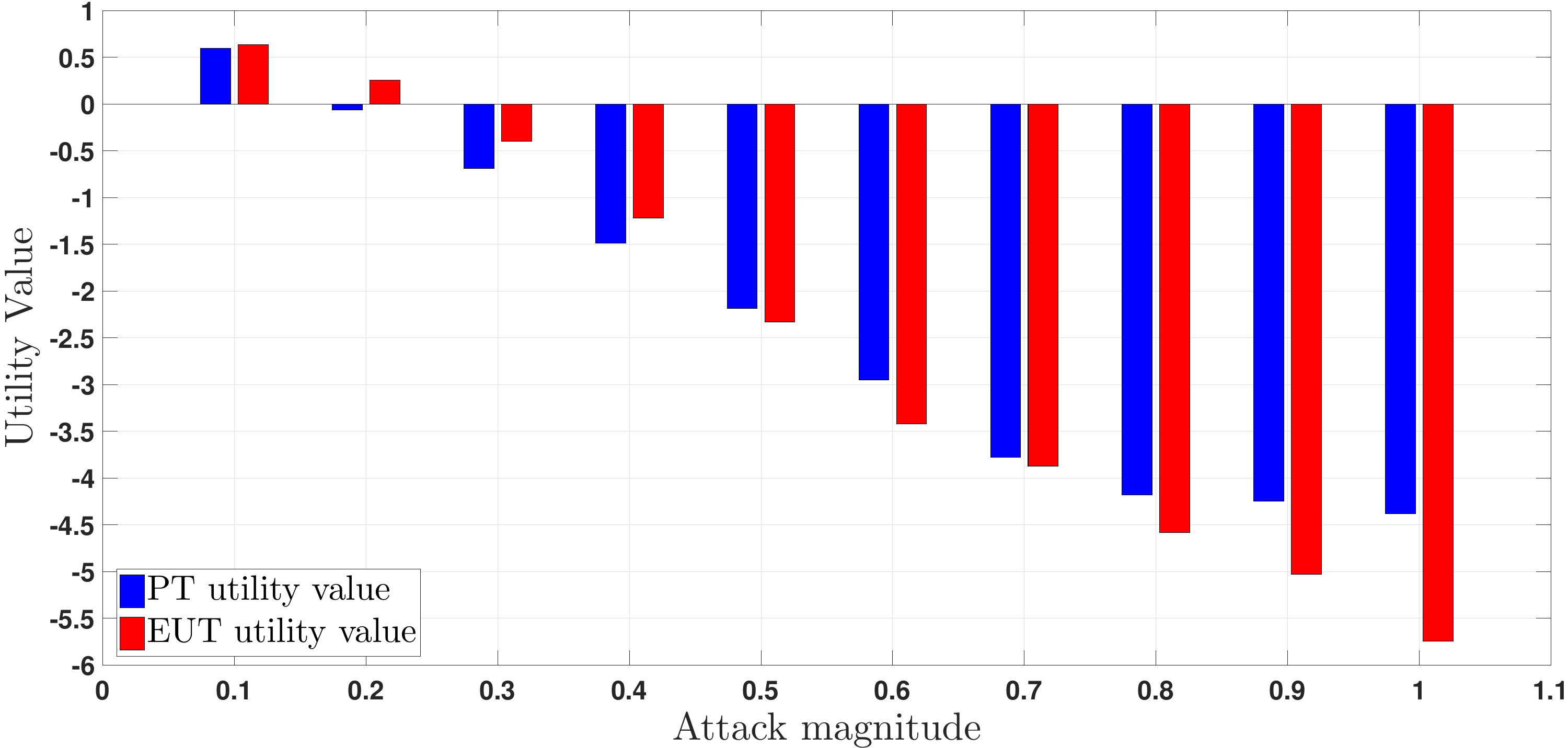}
\caption{Average utility values of an optimistic system after 300 iterations obtained by PT and EUT under different attack magnitudes.}
\label{EUT2}
\end{center}
\end{figure}

We analyzed the effect of expected utility theory on different attack magnitudes in Section~\ref{sec-comparison} and predicted the possible differences between prospect theory and expected utility theory for different attack magnitudes. To validate
those predictions, we have simulated an optimistic system using both prospect theory and expected utility theory. The effect of attack magnitudes on PT and EUT utility values is shown in Figure~\ref{EUT2}. 
The simulation results perfectly match with our predictions.

\begin{figure} [!t]
\begin{center}
\includegraphics[width=3.5in, height=2in,trim=0.15cm 0 0 0cm]{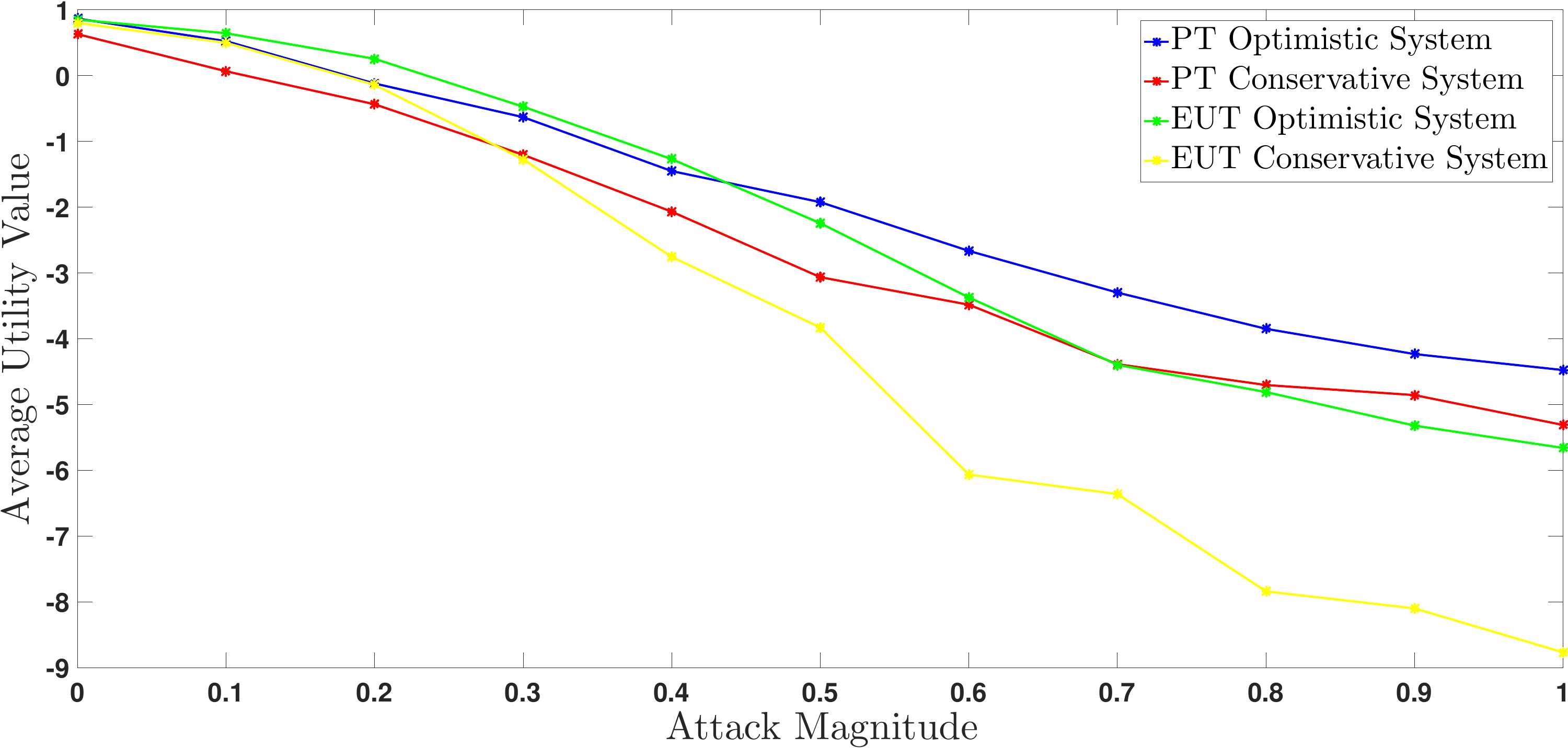}
\caption{Average utility values of optimistic and conservative systems over time vs attack magnitude obtained by PT and EUT.}
\label{EUT3}
\end{center}
\end{figure}

In Section~\ref{sec-comparison}, we also claimed that the effect of expected utility theory on optimistic and conservative systems is different. Expected utility does not use any weighting function so large probabilities have a higher effect on EUT than PT. On the other hand, since malicious data inputs in conservative systems have a higher cost, this phenomenon is more sensible in conservative systems. According to Figure~\ref{EUT3}, the absolute EUT utility values for conservative systems are much higher than absolute PT utility values for conservative systems which aligns with our prediction. This effect is to the extent that even absolute EUT utility values for optimistic systems are larger than absolute PT utility values for conservative systems.

\begin{figure} [!b]
\begin{center}
\includegraphics[width=3.5in, height=2.5in,trim=0.15cm 0 0 0cm]{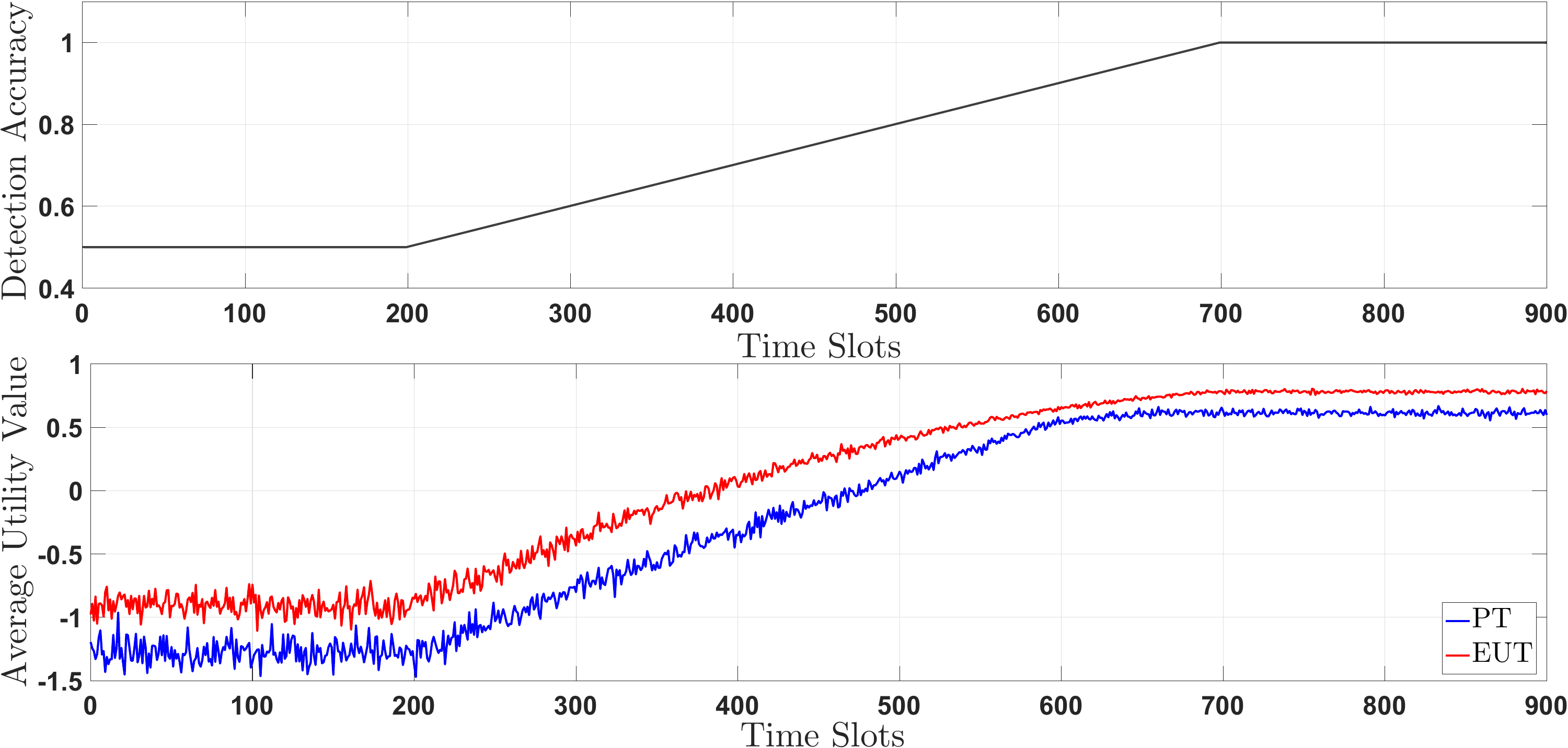}
\caption{Average utility values of an optimistic system over time with increasing $P_{detect}$ obtained by PT and EUT under the same attack magnitude ($P_{attack}$ = 0.1).}
\label{EUT4}
\end{center}
\end{figure}

As was discussed, the three weighting functions associated with probabilities of each outcome in PT contribute to the difference between the utility values obtained by EUT and PT. According to Figure~\ref{EUT4}, under $P_{attack} = 0.1$, the expected utility value is higher than prospect theory utility value when $P_{detect} = 1$ because $P_{attack} = 0.1$ has higher negative effect on PT than EUT due to value and weighting functions. As we reduce the detection accuracy from $P_{detect} = 1$, utility values obtained by PT and EUT start to decrease. However, this negative effect in PT is more than EUT until we reach $P_{detect} = 0.7$ ($P_{undetected}\geq0.3$). According to Figure~\ref{weight function}, in prospect theory, we consider higher weights than the actual probabilities for small probabilities (less than 0.3). Therefore, it is expected that reduction of $P_{detect}$ from 1 to 0.7 which is equivalent to increase of $P_{undecided}$ from 0 to 0.3 has higher negative effect on the utilities obtained by prospect theory than expected utility theory. However, according to Figure~\ref{weight function}, the trend of this effect will change for larger probabilities (larger than 0.3) which means utility values obtained by EUT should decrease with higher rate. However, as the detection accuracy decreases, the number of undecided data inputs increases. According to Figure~\ref{value function}, the increase of undecided data inputs will cause more significant negative effect on PT than EUT. Therefore, as shown in Figure~\ref{EUT4}, for $P_{detect}\leq0.7$ ($P_{undecided}\geq0.3$), these two effects cancel out each other in a fashion that utility values obtained by EUT and PT decrease with the same rate.

\subsection{Trust Management Under On-Off Attacks}
\label{sec-trustmanage}
We consider an adversary launching On-Off attacks in five stages over 500 time slots. 
The first 300 slots are an active attack period and the time range of 300-500 is used to study how trust scores recover. 
For simulations, we consider an Off-On ratio of 2:1. Later, we compare results with 3:1 ratio as well. 
We plot the performance of our proposed method, asymmetric weighted moving average (AWMA), considering different attack scenarios discussed in Section~\ref{sec-AWMA}. Thereafter, we compare the results with other popular trust update schemes and justify the suitability of AWMA with regard to On-Off attacks.

\subsubsection{Apt Weighing Factors and Threshold for AWMA:}
\label{sec-factors}
The weighing factors $\chi_{a}$, $\chi_{b_{max}}$, $\chi_{c_{min}}$, and $\chi_{d}$ are chosen as $0.99$, $0.999$, $0.001$, and $0.001$. We can verify that these factors satisfy the conditions: $0< \chi_{c_{min}} << \chi_{b_{max}} < 1$, $0 < \chi_{a} < 1$, and $ 0< \chi_{d} <1$. The skewed values of the weighing factors $\chi_{c_{min}}$ and $\chi_{b_{max}}$ justify the asymmetry caused by giving a very high weightage on first occurrence of negative behavior and a very low weightage on the first occurrence of positive behavior. The choice of $\chi_{a}$ and $\chi_{d}$ can be used to control the rate of trust redemption. If a system requires slower trust redemption, lower values of $\chi_{a}$ and $\chi_{d}$ are necessary. By considering the four possible cases, discussed in Section~\ref{sec-AWMA}, we update the trust scores. Since there is no fixed attack magnitude in On-Off attacks, we keep the mid-point ($w_{di}(t)=0$) in the weighted trust score range ($w_{di}(t)\approx[-1,+1]$) as the threshold ($\Gamma_{on-off}=0$). However, $\Gamma_{on-off}$ can be adjusted according to the requirements of the system. More conservative systems will have $\Gamma_{on-off}>0$. Also, different values of $\chi_{min}$ and $\chi_{max}$ can be selected to ensure more fairness to nodes in a network susceptible to more bit flips due to noise. 

\subsubsection{AWMA vs CWMA:}
\begin{figure} [!t]
\begin{center}
\includegraphics[width=3.5in, height=2in]{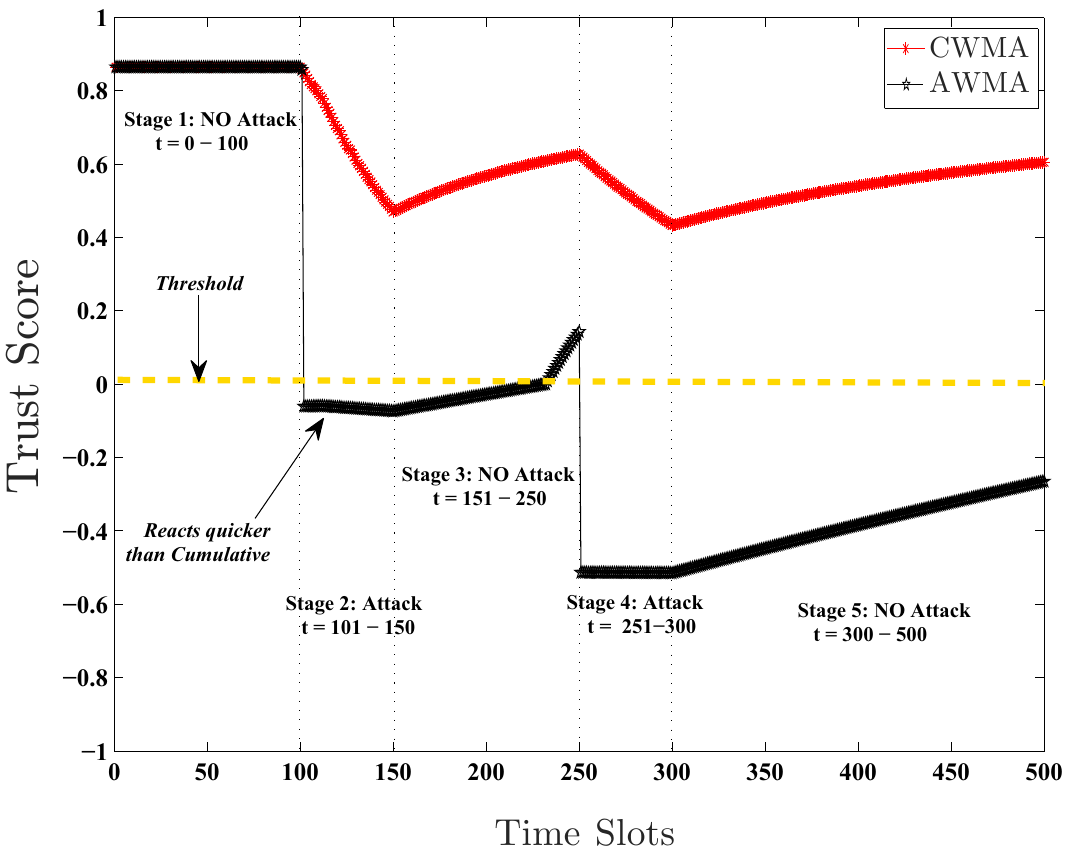} 
\caption{AWMA vs CWMA with 2:1 Off to On Ratio.}
\label{asy_ma_vs_eq_ma}
\end{center}
\end{figure}
In Figure~\ref{asy_ma_vs_eq_ma}, we show how AWMA outperforms CWMA. We observe that at Stage 1 with no attacks (Off mode), both schemes preserve a high trust score. However, when the attack period starts (On mode), from the $101^{th}$ time slot until the next $50$ time slots, AWMA ensures that the cumulative moving average (trust score) is decreased more rapidly and preserves a low value. On the other hand, CWMA slowly reacts to attack because of the good behavior of adversary during the first $100$ time slots. AWMA quickly reacts to attack because once the current weighted trust score ($w_{di}(t)$) drops below zero, it forgets about previous high trust scores ($w_{di}^{mavg(t-1)}$) by using a very small weight ($1-\chi_{b_{max}}=0.001$) for previous trust scores and assigning an extremely high weight ($\chi_{b_{max}}=0.999$) to the current weighted trust score. This aspect happens at the beginning of $101^{th}$ and $251^{th}$ time slots in Figure~\ref{asy_ma_vs_eq_ma}. During recovery phase in Stage 3 and Stage 5 that adversary starts to behave good after an attack, trust score starts increasing in both AWMA and CWMA. However, AWMA does not quickly recover the trust score and takes into account historical malicious behavior. This impact is to the extent that at the end of Stage 5 that attacks have stopped for the last 200 slots, the trust score is still low enough to reflect the adversary's historical malicious behavior. On the contrary,  the cumulative weighted moving average (CWMA) fails to capture the short-term attacks for future decisions. This behavioral difference occurs because in AWMA, previous cumulative moving average of less than $0$ (selected $\Gamma_{on-off}$) is given a higher weight compared to current honest behavior which prevents the trust score to improve very fast during honest behavior.  

\begin{figure} [!t]
\begin{center}
\includegraphics[width=3.5in, height=2in]{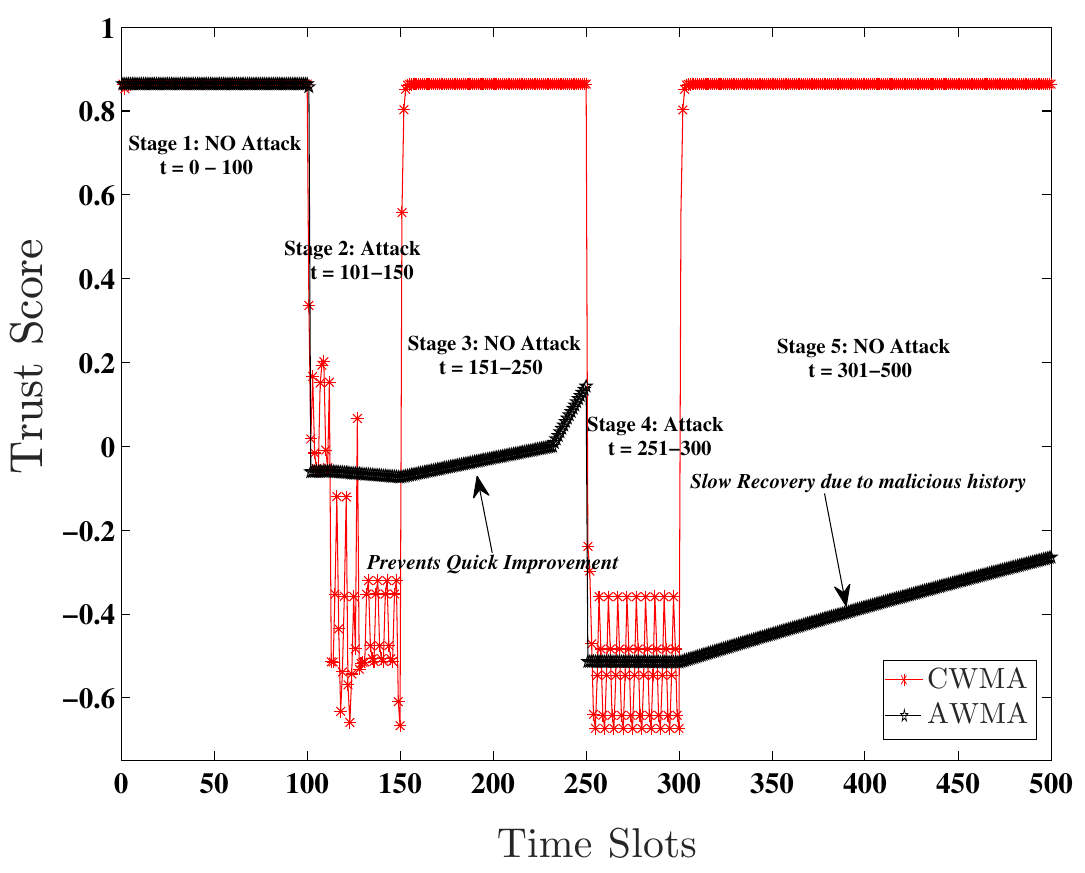} 
\caption{AWMA vs EWMA with 2:1 Off to On Ratio.}
\label{asy_ma_vs_exp_ma}
\end{center}
\end{figure}

\subsubsection{AWMA vs EWMA:}
EWMA can quickly react when an attack is initiated. However, when the malicious node starts to behave honestly, it forgets node's malicious behavior as quickly as its reaction to attack.
This behavior is unfavorable because the system should not be allowed to redeem its trust score quickly unless it experiences a long period of honest behavior. The case (c) of AWMA scheme, discussed in Section~\ref{sec-AWMA}, addresses this deficiency where we assign a very small weight to honest behavior pursued by a period of dishonest behavior. Hence, trust score in AWMA slowly increases. In Figure~\ref{asy_ma_vs_exp_ma}, we do not see much difference at stage 1 due to no attack. Also, there is not much difference at Stage 2 since large weight is given to new trust scores by both schemes. However, at Stage 3, EWMA allows the malicious data input to quickly recover its trust score due to forgetting old malicious behavior. On the other hand, AWMA does not forget previous low trust scores caused by malicious behavior. This is realized at the beginning of Stage 3 by assigning larger weights to previous cumulative moving average value compared to current honest behavior. This prevents fast improvement of the trust scores even during the period of honest behavior. We see that for all subsequent stages the exponentially weighted averages oscillate between high and low scores, but AWMA preserves a low score while maintaining fairness by allowing a very slow increase of cumulative trust score at stage 5 due to its continuous good behavior for 200 slots. 

\subsubsection{Effect of Off-On Ratio:}
\begin{figure} [h]
\begin{center}
\includegraphics[width=3.5in, height=2.3in]{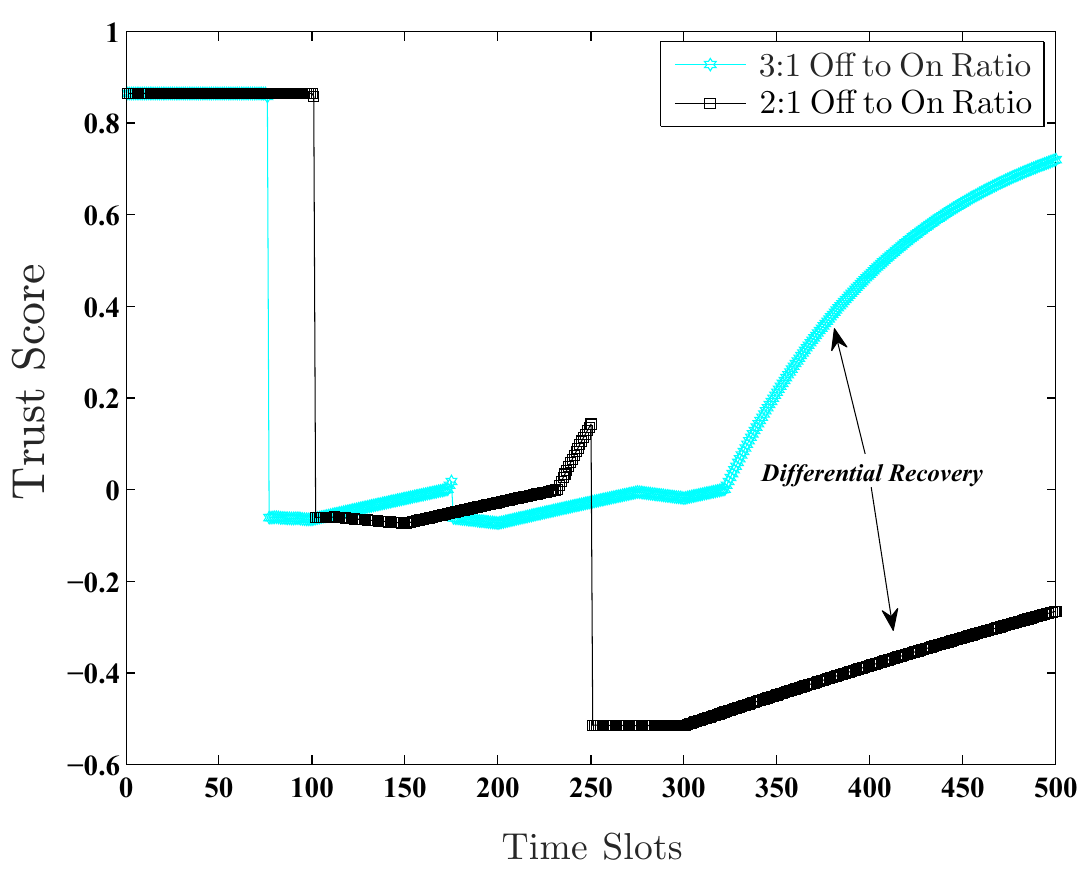} 
\caption{Comparison between 2:1 and 3:1 attack ratios.}
\label{compare_awmaz}
\end{center}
\end{figure}

The Off-On attack ratio of 3:1 is less aggressive than 2:1. Hence, after 500 slots, we should expect 3:1 to have higher trust score. It should be noted that a ratio as high as 1:1 is not a characteristic of On-Off attacks. Also, too low attack ratios hardly can affect a system. In Figure~\ref{compare_awmaz}, we observe the differences in the trust scores under 2:1 and 3:1 attack ratios.

\section{Conclusions and Future work}
In this paper, we surveyed the deficiencies of existing methods for trust management in IoT networks. We explained there is a need for an approach to be adaptive to opportunistic (On-Off) attacks, consider an imperfect anomaly monitoring mechanism, and differentiate between systems with different risk tolerance. Therefore, we proposed a prospect theoretic approach to evaluate the data integrity in an IoT network that is exposed to 
opportunistic and non-opportunistic data manipulation attacks. By considering an imperfect anomaly monitoring mechanism, we quantified the trustworthiness of the aggregate data at an IoT hub using utility values obtained by prospect theory and expected utility theory. 
A comparison was drawn between these two theories considering their features and their applicability 
to trust scoring in an IoT network. According to the theoretical predictions and simulation experiments, 
prospect theory proved to be more promising for measuring trustworthiness of the aggregate data 
in risk-averse systems. Eventually, we put forward  asymmetric weighted moving average (AWMA) scheme as an extension of our approach to take into account stealthy On-Off attacks for trust management. 
The proposed framework was validated using extensive simulation experiments and the results bring out the efficacy 
of the proposed framework.

Although our proposed method solves several limitations of existing trust management systems for IoT networks, it suffers from centralization and being susceptible to single point of failure. In future, we will address this shortcoming by applying blockchain solutions. Blockchain has proven to be very promising for decentralization of IoT networks~\cite{salimitari2018survey, roy2019cache}. In addition, AI-enabled blockchain platforms can significantly improve fault tolerance and security of IoT networks~\cite{salimitari2019ai}.

\appendix
\section{APPENDICES}
\subsection{Equation~(\ref{eq14})}
\label{ap1}
For simplicity, let us define the following variable from Equation~(\ref{eq11}): 
\begin{equation}
I_1 = \!\!\!\!\!\!\!\!\!\!\!\!\!\!\!\!\!\!\!\!\!\!\!\!\!\!\! \int\limits_{\qquad\qquad\quad D(N)(\theta_{\alpha},\theta_{\beta},\theta_{\mu})} \!\!\!\!\!\!\!\!\!\!\!\!\!\!\!\!\!\!\!\!\!\!\!\!\!\!\!\!\! \theta_{\alpha}^{n_{\alpha}}\theta_{\beta}^{n_{\beta}}\theta_{\mu}^{n_{\mu}}d\theta_{\alpha}d\theta_{\beta}d\theta_{\mu}\raisepunct{,}
\label{eq21}
\end{equation}
where, $I_1$ can be solved by using multivariate generalization of the Eulerian integral of first kind. Note that ${D(N)(\theta_{\alpha},\theta_{\beta},\theta_{\mu})}$ denotes a space and we know that a space of $m$ (in this case, $m=3$) parameters has only $m-1$ degrees of freedom
 due to the additivity constraint $\theta_{\alpha}+\theta_{\beta}+\theta_{\mu}=1$. Therefore, when we integrate over this space, the integration has $m-1=2$ dimensions. Hence, $I_1$ can be written as:
\begin{equation}
I_1=\int_{0}^{1}\!\!\!\int_{0}^{1-\theta_{\alpha}-\theta_{\beta}}\!\!\theta_{\alpha}^{(n_{\alpha}+1)-1}\theta_{\beta}^{(n_{\beta}+1)-1} (1-\theta_{\alpha}-\theta_{\beta})^{(n_{\mu}+1)-1}d\theta_{\alpha}d\theta_{\beta}\raisepunct{.}
\label{eq12}
\end{equation}
Equation~(\ref{eq12}) is a known form for the multivariate extension of the Beta function which in this case is defined as $B(n_{\alpha}+1,n_{\beta}+1,n_{\mu}+1).$
The proof can be found in Lemma 2.4.1 of~\cite{Fang}.
In general:
\begin{eqnarray}
B(\alpha_1,\,\dots\,,\alpha_m)
&=& \!\!\!\!\!\!\!\!\!\!\!\!\!\!\!\!\!\!\!\!\!\!\!\!\!\!\!\int\limits_{\qquad\qquad\quad D(x_1,\,\dots\;,x_{m-1})}\!\!\!\!\!\!\!\!\!\!\!\!\!\!\!\!\!\!\!\!\!\!\!\!\!\!\!\!\! x_{1}^{\alpha_{1}-1}\;\dotsm\;(1-\sum_{i=1}^{m-1}x_i)^{\alpha_{m}-1}dx_1\dotsm\, dx_{m-1}\nonumber\\
&=& \frac{\prod_{i=1}^{m}\Gamma(\alpha_{i})}{\Gamma(\sum_{i=1}^{m}\alpha_{i})} = \frac{\Gamma(\alpha_1)\dotsm\Gamma(\alpha_m)}{\Gamma(\alpha_{1}+\dotsm+\alpha_m)}\raisepunct{.}
\label{13a}
\end{eqnarray}
Using the above result, we can rewrite Equation~(\ref{eq12}) as:
\begin{eqnarray}
I_1 &=& B(n_{\alpha}+1,n_{\beta}+1,n_{\mu}+1)  \nonumber \\
&=&\frac{\Gamma(n_{\alpha}+1)\Gamma(n_{\beta}+1)\Gamma(n_{\mu}+1)}{\Gamma(n_{\alpha}+1+n_{\beta}+1+n_{\mu}+1)}\raisepunct{.}
\label{13b}
\end{eqnarray}
Putting Equation~(\ref{13b}) in Equation~(\ref{eq11}), we get:
\begin{equation}
P(D(N))=\frac{N!}{n_{\alpha}! n_{\beta}! n_{\mu}!} \frac{\Gamma(n_{\alpha}+1)\Gamma(n_{\beta}+1)\Gamma(n_{\mu}+1)}{\Gamma(n_{\alpha}+1+n_{\beta}+1+n_{\mu}+1)}\raisepunct{.}
\label{eq13}
\end{equation}
Since the parameters in gamma functions ($n_{\alpha}+1$, $n_{\beta}+1$, and so on) are all non-zero positive values, we can use the property that $\Gamma(z)=(z-1)!$ to rewrite Equation~(\ref{eq13}) as:
\begin{equation}
P(D(N))=\frac{N!}{n_{\alpha}! n_{\beta}! n_{\mu}!} \frac{n_{\alpha}!n_{\beta}!n_{\mu}!}{(N+2)!}\raisepunct{.}
\label{eq22}
\end{equation}
By cancelling out the parameters, Equation~(\ref{eq14}) is obtained.

\subsection{Equation~(\ref{eq17})}
\label{ap2}
Assuming conditional
independence between $\hat{X}(\bar{\theta})$, $D(N)$ and $\bar{\theta}$,
we calculate the numerator of Equation~(\ref{eq9}) as:
\begin{eqnarray}
P(\hat{X}(\bar{\theta})=\alpha,D(N))
& = & \!\!\!\!\!\!\!\!\!\!\! \int\limits_{\qquad D(N)(\bar{\theta})}\!\!\!\!\!\!\!\!\!\!\!\!P(X(\bar{\theta})=\alpha,D(N)| \bar{\theta})f(\bar{\theta})d(\bar{\theta})\nonumber\\
& = & \!\!\!\!\!\!\!\!\!\!\! \int\limits_{\qquad D(N)(\bar{\theta})}\!\!\!\!\!\!\!\!\!\!\!\!P(X(\bar{\theta})=\alpha| \bar{\theta})P(D(N)|\bar{\theta})d(\bar{\theta})\nonumber\\
& = &\frac{N!}{n_{\alpha}! n_{\beta}! n_{\mu}!}\!\!\!\!\!\!\!\!\!\!\!\!\!\!\!\!\!\!\!\!\!\!\!\!\!\!\!\int\limits_{\qquad\qquad\quad D(N)(\theta_{\alpha},\theta_{\beta},\theta_{\mu})}\!\!\!\!\!\!\!\!\!\!\!\!\!\!\!\!\!\!\!\!\!\!\!\!\!\!\!\!\!\theta_{\alpha}\theta_{\alpha}^{n_{\alpha}}\theta_{\beta}^{n_{\beta}}\theta_{\mu}^{n_{\mu}}d\theta_{\alpha}d\theta_{\beta}d\theta_{\mu}\nonumber\\
& = & \frac{N!}{n_{\alpha}! n_{\beta}! n_{\mu}!}\!\!\!\!\!\!\!\!\!\!\!\!\!\!\!\!\!\!\!\!\!\!\!\!\!\!\!\int\limits_{\qquad\qquad\quad  D(N)(\theta_{\alpha},\theta_{\beta},\theta_{\mu})} \!\!\!\!\!\!\!\!\!\!\!\!\!\!\!\!\!\!\!\!\!\!\!\!\!\!\!\!\!
\theta_{\alpha}^{n_{\alpha}+1}\theta_{\beta}^{n_{\beta}}\theta_{\mu}^{n_{\mu}}d\theta_{\alpha}d\theta_{\beta}d\theta_{\mu}\raisepunct{.}
\label{eq15}
\end{eqnarray}
The above integral has the same form as Equation~(\ref{eq21}).
Hence, the integral portion of Equation~(\ref{eq15}), named $I_2$, can be rewritten similar to Equation~(\ref{eq12}) as:
\begin{equation}
I_2= \int_{0}^{1}\!\!\!\int_{0}^{1-\theta_{\alpha}-\theta_{\beta}}\!\!\theta_{\alpha}^{(n_{\alpha}+2)-1}\theta_{\beta}^{(n_{\beta}+1)-1} (1-\theta_{\alpha}-\theta_{\beta})^{(n_{\mu}+1)-1}d\theta_{\alpha}d\theta_{\beta}\raisepunct{.}
\label{eq16}
\end{equation}
After solving $I_2$ by using Equation~(\ref{13a}), we can update Equation~(\ref{eq16}) as:
\begin{equation}
P(\hat{X}(\bar{\theta})=\alpha,D(N)) = \frac{N!}{n_{\alpha}! n_{\beta}! n_{\mu}!} \frac{(n_{\alpha}+1)!n_{\beta}!n_{\mu}!}{(N+3)!}\raisepunct{.}
\label{eq23}
\end{equation}
By simplifying Equation~(\ref{eq23}), we obtain Equation~(\ref{eq17}).

\end{document}